\documentstyle[a4,12pt,epsfig]{article}

\begin{document}
\renewcommand{\thefootnote}{\fnsymbol{footnote}}
\newcommand{\as}{\mbox{$\alpha_{\mathrm{s}}$}}
\newcommand{\xmu}{\mbox{$x_{\mu}$}}
\newcommand{\amz}{\mbox{$\alpha_{\mathrm{s}}(\mathrm{M}_{\Zzero})$}}
\newcommand{\oaa}{\mbox{$\cal{O}(\alpha_{\mathrm{s}}^2$})}
\newcommand{\epem}{\mbox{$\mathrm{e^+e^-}$}}
\newcommand{\Zzero}{\mbox{${\mathrm{Z}^0}$}}
\newcommand{\WW}{\mbox{$\mathrm{W^+W^-}$}}
\newcommand{\qq}{\mbox{$\mathrm{q\overline{q}}$}}
\newcommand{\ppbar}{\mbox{$\mathrm{p\overline{p}}$}}
\newcommand{\Zqq}{\mbox{$ \Zzero / \gamma \rightarrow \qq $}}
\newcommand{\WWqqqq}{\mbox{\WW$\rightarrow$\qq\qq}}
\newcommand{\WWqqln}{\mbox{\WW$\rightarrow$\qq$\ell\overline{\nu_{\ell}}$}}
\newcommand{\mz}{\mbox{$M_{\mathrm{Z}^0}$}}
\newcommand{\mw}{\mbox{$M_{\mathrm{W}}$}}
\newcommand{\ecm}{\mbox{$E_{cm}$}}
\newcommand{\Opal}{\mbox{O{\sc pal}}}
\newcommand{\LepII}{\mbox{L{\sc ep}II}}
\newcommand{\LepI}{\mbox{L{\sc ep}I}}
\newcommand{\Jetset}{\mbox{J{\sc etset}}}
\newcommand{\Pythia}{\mbox{P{\sc ythia}}}
\newcommand{\Herwig}{\mbox{H{\sc erwig}}}
\newcommand{\half}{\mbox{$\textstyle\frac{1}{2}$}}
\newcommand{\boldp}{\mbox{\boldmath$p$}}
\newcommand{\boldn}{\mbox{\boldmath$n$}}
\newcommand{\Mbar}{\mbox{$\overline{M}$}}
\newcommand{\costh}{\mbox{$\cos\theta$}}
\newcommand{\sprime}{\mbox{$s^{\prime}$}}
\begin{titlepage}
%
\bigskip
\begin{tabbing}
\` \today \\
\end{tabbing}
\begin{center}
{\LARGE\bf Computation of confidence levels
for exclusion or discovery of a signal with the method
of fractional event counting
 \\}
\bigskip P.Bock \\
Physikalisches Institut der Universit\"at Heidelberg, Germany
\end{center}\bigskip

\begin{center}{\Large\bf Abstract}

A method is described, which computes from an observed
sample of events upper limits for production rates of 
particles, or, in case of appearance of a signal,
the probability for an upwards fluctuation of the background.
For any candidate, a weight is defined, and the 
computation is based on the sum of observed weights.
Candidates may be distributed over many decay channels 
with different detection efficiencies, physical observables and 
different or poorly known background. Systematic errors with
any possible correlations are taken into account and they are 
incorporated into the weight definition. It is investigated, 
under which conditions a Bayesian treatment of systematic errors is 
correct. Some numerical examples are given and compared with 
the results of other methods.
Simple approximate formulas for observed and expected confidence 
levels are given for the limiting case of high count rates. 
A special procedure is introduced, which analyses input 
data in terms of polynomial distributions. It extracts confidence
levels for a signal or background hypothesis on the basis of
spectral shapes only, normalizing the total rate to the number of
observed events.
\end{center}
\noindent
 
\end{titlepage}

\section{Introduction}

The operation of high energy accelerators like the LEP storage ring, 
HERA or the Tevatron opened the field of searches for new particles 
beyond the standard model. The search for the last missing standard
model particle, the Higgs boson, was extended up to a mass of 114 GeV at 
LEP \cite{LEP03}. 
The analyses are often quite complex,
because many physical channels have to be combined and sophisticated 
and efficient event taggers have been developed to find certain
event topologies. In case of a data excess over expected background,
the question comes up immediately, whether an upwards fluctuation of 
the background can be ruled out or an event excess can be 
attributed to the particle which is searched for.

A lot of literature exists which adresses these questions
(see the summaries given in refs. \cite{workshop1} to 
\cite{workshop3}. Many publications refer to simple counting
experiments.

In this paper, the method of fractional event counting is described,
which uses a weighted sum over the observed events  as the indicator 
for a signal. The weights (or filter function) are extracted
from physical variables of the candidates, and they 
have to be defined to use the experimental information in a 
statistically optimal way. The weight optimization is done without
use of observed data and is based on Monte Carlo predictions for the
signal and background. The event weighting allows it to avoid hard
cuts in event accpetance which may be subjective: precuts can be
placed in phase space regions where the weight is very small.

This paper summarises the current status of the method, because it has 
been used in some analyses (refs. \cite{opal1} to \cite{LEP99}).
Recently, the sensitivity of the method was improved by incorparating
systematic errors in the filter function. In the past, systematic
errors were finally folded in, but the basic candidate 
weights were defined without it.

The statistical analysis uses the frequentist approach.
Bayesian statistics has been applied in a similar way to the 
multichannel case too \cite{obraz}, and there are even 
comparative results for one physical analysis \cite{LEP99}.

\section{Specification of the filter function}
\subsection{Discriminating variables}
The aim of any statistical analysis of a search experiment is the 
distinction between two physical hypotheses:
\begin{itemize}
\item (A) The data consist of background and the physical signal.
\item (B) The data consist of background only.
\end{itemize}
A discriminating variable, $\xi$, is introduced to
order observed events according to their signal likeness.
This variable can be the particle mass in the search for a 
resonance, a likelihood constructed from some physical observables
or the output variable of a neural network. It is assumed that
theoretical predictions for the spectral distributions of signal 
and background, $s(\xi)$ and $b(\xi)$, exist. 

Data may be available for more than one decay mode of a particle and 
searches may be performed at several acclerator
energies by more than one experiment. All these results have to be
combined. The data will therefore be ordered according to search 
channels. The $\xi$ variable will vary from channel to channel.

All searches are assumed to be statistically independent.
It is therefore never allowed that the same event appears twice.
If an overlap exists, for instance between two final states looked at,
the two corresponding channels must be rearranged into
three: exclusive selection of events in the two original channels 
and the overlap between the two with a new definition of $\xi$. 

In most cases the signal and background spectra of $\xi$ will be 
available in form of Monte Carlo histograms $s_{ki}=s(\xi_{ki})$ and 
$b_{ki}=b(\xi_{ki})$. Here, the index $k$ is used to identify
a channel and $i$ indicates the value of its discriminating 
variable. The trivial case of event counting corresponds to the 
limitation to one histogram bin. Throughout this paper it is assumed
that histograms are normalized to the expected rates, any bin contains
the local mean rate. 
It may be that the total signal rate $r=\sum_{ki} s_{ki}$ has to be 
varied during the analysis.
For later convienience, signal efficiencies per bin can be defined as
\begin{eqnarray}
\epsilon_{ki} = \frac{s_{ki}}{r} \nonumber
\end{eqnarray}
Branching ratios of decays,
channel dependent cross sections and different luminosities are 
incorporated into the $\epsilon_{ki}$ definition.

If a likelihood or neural network definition does not contain 
the particle mass explicitly, but a reconstructed mass exists and is 
not correlated strongly to the likelihood, the mass and 
likelihood distributions $D(m)$ and $D(L)$ can be combined to 
define $\xi$. This will be described in section \ref{w2}.

Instead of any $\xi$, a monotone function of it can be used as
discriminating variable too. Apart from binning effects,
the final results will be independent of such a redefinition.
This will be shown in the next subsection.  
The choice of $\xi$ is rather arbitrary and has to be based on 
physical arguments and numerical convenience.

\subsection{Event weights}
\label{evw}
From $s_{ki}$ and $b_{ki}$, event weights $w_{ki}$ will be computed.
The definition of $w_{ki}$ is not unique. Every new filter gives
another result for the same experiment, and all procedures are 
correct on statistical average. However, different definitions do
not have the same performance and the filter should be optimized to 
get the best separation between hypotheses (A) and (B).

If the $w_{ki}$ are known, the total weight of an event sample, 
often called 'test statistics' X, is defined as
\begin{eqnarray}
X = \sum_l w_{k(l)i(l)} \nonumber
\end{eqnarray}
The sum extends over all candidates of an experimental data set or
a Gedanken experiment. The indices $k(l)$ indicate the channels
and $i(l)$ are the $\xi$ bins to which the events belong.

If an experiment would be repeated many times, the resulting total 
weights show statistical fluctuations. They have to be described with
probability density functions $P_b(X)$ and $P_{sb}(X)$. These
functions refer to the hypotheses (B) (background only) and 
(A) (the signal exists). They are related to the input histograms
$s_{ki}$ and $b_{ki}$ and depend on the filter specification.
Implicitly they depend on the total signal and background rates.
Their computation will be described in detail 
in the next section.

Confidence levels for a background or a signal plus background
compatibility of a special data set with the test statistics 
$X=W_{tot}$ can be computed with
\begin{equation}
\label{CLdef}
CL_b(W_{tot})=\int_0^{W_{tot}} P_b(X)  dX;  \hspace*{7mm}
CL_{sb}(W_{tot})=\int_0^{W_{tot}} P_{sb}(X) dX
\end{equation} 
These definitions are based on the frequentist approach.
If a hypothesis is true, the median value of the corresponding
confidence level is 1/2. A small value of $CL_{sb}$ indicates a 
data deficit, if hypothesis (A) is true: $CL_{sb}$ is the 
frequency of a downward fluctuation of $X$ at least to $W_{obs}$, 
and somewhat unprecisely it is said that 
hypothesis (A) is ruled out with probability $1-CL_{sb}$. 
Vice versa, if $CL_b$ is close to 1, a data
excess over background is observed which will appear with frequency
$1-CL_b$, if no signal exists.

It is now straight forward to optimize the definition of $w_{ki}$ 
in the limit of high rates. According to the central limit theorem
the functions $P_{sb}$ and $P_b$ have approximately Gaussian shape:
\begin{eqnarray}
 P_b(X)=\frac{1}{\sqrt{2 \pi} \sigma_b} \exp (-\frac{(X-<X>_b)^2}
         {2 \sigma_b^2}) \nonumber \\
 P_{sb}(X)=\frac{1}{\sqrt{2 \pi} \sigma_{sb}} \exp (-\frac{(X-<X>_{sb})^2}
         {2 \sigma_{sb}^2}) \label{xgaus}
\end{eqnarray}
The expectation values of $X$ are given by
\begin{equation}
\label{meandef}
<X>_b = \sum_{k,i} w_{ki} b_{ki}  \hspace*{5mm} 
<X>_{sb} = <X>_s + <X>_b = \sum_{k,i} w_{ki} (r \epsilon_{ki} + b_{ki})
\end{equation}
The sums extend over all channels and $\xi$ bins.
The statistical errors introduce the variances
\begin{equation}
\label{vardef}
\sigma^2_b = \sum_{k,i} w_{ki}^2 b_{ki}  \hspace*{5mm} 
\sigma^2_{sb} = \sigma^2_s + \sigma^2_b 
  = \sum_{k,i} w_{ki}^2 (r \epsilon_{ki} + b_{ki})
\end{equation}

Criteria for optimal discrimination between hypotheses (A) and (B) are:
\begin{itemize}
\item (i) The mean confidence level for interpretation of an arbitrary
test statistics $X$ from the background source (B) as signal
plus background (A),\newline $<CL_{sb}>_b$, should be a minimum.
\item (ii) The mean confidence level for interpretation of an arbitrary
test statistics $X$ from the combined signal and background source
(A) as background (B), $<CL_b>_{sb}$, should be a maximum.
\end{itemize}
The first confidence level is simply the mean probability that an
arbitrary Gedanken experiment with signal and background events 
has a total weight less than or equal to the weight of an 
arbitrary experiment counting background. The second confidence 
level is complementary so that both optimization criteria are 
identical.

In the high rate limit 
the probability densities at $X=0$ are negligible and one gets
with equations (\ref{CLdef}) to (\ref{vardef}): \newpage
\begin{eqnarray}
<CL_{sb}>_b=\frac{1}{\sqrt{2 \pi} \sigma_b} \int_{-\infty}^{\infty}
dX \cdot \exp (-\frac{(X-<X>_b)^2}{2 \sigma^2_b}) \cdot \nonumber \\ 
 \frac{1}{\sqrt{2 \pi \cdot (\sigma_b^2 + \sigma^2_{sb})}}
 \int_{-\infty}^X dY \cdot \exp (-\frac{(Y-<X>_{sb})^2}{2 \sigma^2_{sb}} )
 \nonumber
\end{eqnarray}
On the left hand side, the following conventions are introduced:
The brackets indicate the statistical mean value.
Both physical models appear in the equation.
The events consist of background, which is indicated by the index
'$b$' on the left hand side, 
but they are analysed by the observer in terms of signal and
background ($CL_{sb}$). 
The double integral can be simplified to
\begin{eqnarray}
\label{aclsb}
<CL_{sb}>_b= \frac{1}{\sqrt{2 \pi \cdot (\sigma^2_{sb} + 2 \sigma_b^2)}}
 \int_{-\infty}^{-<X>_s} dZ \cdot \exp (-\frac{Z^2}
                    {2 (\sigma^2_{sb} + 2 \sigma_b^2)})
\end{eqnarray}
where
\begin{eqnarray}
<X>_s = \sum_{k,i} w_{ki} s_{ki} \nonumber
\end{eqnarray}
is the expectation value of $X$ for signal events.
The probability $<CL_{sb}>_b$ depends on the ratio 
$ <X>_s/\sqrt{\sigma^2_{sb} + 2 \sigma_b^2}$ only which has to be maximized.
Because a common scale factor in all $w_{ki}$ 
cancels out in the confidence levels, the mean value $<X>_s$ can be fixed.
The optimization criterion is then, with a Lagrangian factor $\lambda$,
\begin{eqnarray}
\frac{\partial (\sigma^2_{sb} + 2 \sigma_b^2)}{\partial w_{ki}}
           - \lambda \frac{\partial <X>_s}{\partial w_{ki}} = 0
\nonumber 
\end{eqnarray}
After multiplication with a common constant factor the result becomes simply
\begin{eqnarray}
\label{wdef}
w_{ki} = \frac{s_{ki}}{s_{ki} + 2 b_{ki}}
\end{eqnarray}
The factor 2 appears because the width of the background distribution
enters twice. The weight for a specific channel and bin is independent of
the use of any other channel or bin.

Equations (\ref{wdef}) and (\ref{CLdef}) to (\ref{vardef})
are sufficient to compute confidence levels for a data set in the
high rate approximation.

The above optimization is not unique. There are other possibilities:
\begin{itemize}
\item (iii) One may look for a bound on a predicted 
rate $r$ at a requested confidence level $CL$.
A fixed $CL$ is equivalent to a cut in the signal plus 
background distribution of $X$ at
$X_{cut} = 
r \cdot \sum_{k,i} \epsilon_{ki} w_{ki} + <X>_b - K \cdot \sigma_{sb}$,  
where $K$ is the number of standard deviations equivalent to
$CL$. The probability for an upwards fluctuation of the background 
above the cut $X_{cut}$ should be a minimum.
In the high rate limit the expected value of this probability is
\begin{equation}
\label{eclb}
1- E[CL_b]_{sb} = 1- \frac{1}{\sqrt{2 \pi}\sigma_b} \int_{-\infty}^{X_{cut}}
 \exp ( -\frac{(X-<X>_{b})^2}{2 \sigma_b^2}
\end{equation}
Countrary to eq.(\ref{aclsb}), a sample of signal and background
events is analysed in terms of background. The confidence level
$CL_b$ is not averaged over the whole signal plus background
distribution, it is computed at a fractile of it, which is related
to $K$. 
One gets the condition
\begin{eqnarray}
\frac{( r \sum_{k,} \epsilon_{ki} w_{ki}  - K \sigma_{sb})^2}{\sigma_b^2}
 = \hspace*{2mm} {\rm{max.}} \nonumber
\end{eqnarray}
\item (iv) One can optimize the chance to find a signal, which
exceeds the background prediction at a requested confidence level 
$CL$. This confidence level corresponds to a cut in the 
weight distribution at $X_{cut} = \sum_{k,i} w_{ki} b_{ki} + K \sigma_b$. 
The maximum chance to detect a signal is obtained by minimizing
the probability for a downward fluctuation below $X_{cut}$:
\begin{eqnarray}
\label{eclsb}
E[CL_{sb}]_b = \frac{1}{\sqrt{2 \pi} \sigma_{sb}} \int_{-\infty}^{X_{cut}}
 \exp ( -\frac{(X-<X>_{sb})^2}{2 \sigma_{sb}^2} \\
\frac{(\sum_{k,i} r \epsilon_{ki} w_{ki}  - K \sigma_{b})^2}{\sigma_{sb}^2}
 = \hspace*{2mm} {\rm{max.}} \nonumber
\end{eqnarray}
\item (v) The measurement of a hypothetical signal rate is most
 significant, if the ratio  
\begin{eqnarray}
<X>_s^2 / \sigma^2_{sb} \nonumber
\end{eqnarray}
is maximal. This request corresponds to the special case $K=0$ 
in item (iv): the probability that the total weight of signal and 
background events exceeds the median background level is maximized. 
\end{itemize}

The functional form of $w$ is obtained in the same way as eq.(\ref{wdef}).
After requesting a fixed sum  $\sum_{k,i} w_{ki} \epsilon_{ki}$ to
set the $w_k$ scale, computation of the derivatives with with 
respect to $w_{ki}$, absorption of all $k$ and $i$ independent sums 
into common constants and a final renormalization one finds in any
case the functional form
\begin{equation}
\label{wrdef}
w_{ki} = \frac{{\mathcal{R}} \epsilon_{ki}}{R \epsilon_{ki} 
+ b_{ki}}
\end{equation}
This general result contains a free rate parameter $R$, which has to be
tuned to fulfill a specific optimization criterion (i) to (v).
The confidence levels are invariant against multiplication of all 
$w_{ki}$ with a common factor. For definiteness, the normalization
constant ${\mathcal{R}}$ is introduced to adjust the the overall
maximum weight to 1, but this factor could also be dropped. 
To garantee a positive denominator in any case, $R$ should
be positive.

In general, $R$ is not equal to the signal rate $r$, but it is 
proportional to it: $ R = c \cdot r$.
For condition (v) (best rate measurement) one gets $R = r$. 
In the special case (iii) with $K=0$, the observation of a signal at
its median value and minimum upwards fluctuation of the background,
the result is $R=0$, which means that the weight is proportional to
the signal to background ratio.
Equation (\ref{wdef}) is contained as the special case $R = 1/2 r$, 
which is a good compromise. 

If no signal exists at all, an observer will try to find the lowest
possible upper limit $n_{CL}$ for it. Requesting a definite number 
of standard deviations $K$ for the limit and assuming that only
background is observed at its median level, the observer has 
to solve the equation
\begin{eqnarray}
\sum_{k,i} n_{CL} \epsilon_{ki} w_{ki} - K \sigma_{sb} = 0
\nonumber
\end{eqnarray}
The error  $\sigma_{sb}$ depends on the expected limit $n_{CL}$.
Derivation of the last equation with respect to $w_{ki}$ and setting
$d n_{CL}/d w_{ki} = 0$ gives eq.(\ref{wrdef}) with $R=n_{CL}$.
This is a self consistency relation between the expected rate limit
and the parameter $R$, which depends on $K$.

Solution (\ref{wrdef}) depends on the $\epsilon_{ki}$ to $b_{ki}$
ratio only and is therefore invariant against $\xi$ transformations, 
which rescale both distributions with the same \newline ($\xi$ dependent)
factor.

It was derived in the high rate limit, but
can be applied at low rates too. In this region it is not expected
to be optimal anymore but it is still very close to the optimum and it 
gives still bias free results. Of course, the simple analytic formulas for
the confidence integrals and the results for the $R$ values given here 
will break down.

Throughout this paper it is the understanding that the weight
algorithm, including the parameter $R$, is fixed a priory 
and not fitted to observed data. This makes it necessary to generalize
the criteria (i) to (v) to non-Gaussian distributions and to
compute the functions $P_{sb}, P_b$ and the expected confidence levels
$<CL_{sb}>_b, E[CL_{sb}]_b$ and $E[CL_b]_{sb}$ numerically, 
using theoretical predictions for $\epsilon_{ki}$, $b_{ki}$ and $r$.
The parameter $R$ has to be varied until a requested
optimization criterium is fulfilled. 
 
This ambiguity in defining the weight function is very confusing.
As will be shown later, the optimization procedure allows variations of $R$ 
within rather wide regions, if little numerical tolerances of the
expected confidence levels are accepted. 
Nevertheless, the result for a specific data set is $R$ dependent. 
At large rates, this effect if often small. However, in low statistics 
experiments the analyses become rather ambigious. On statistical
average all results would be correct, but one has to select one
parameter without introducting subjectivity.

In many cases the signal to background ratio ($R=0$) is the suitable
choice. This is especially true, if some signal is observed, but   
no theoretical prediction for the cross section exists. An expected
signal rate is not needed to define $w$ and the function $P_b$ can be
used to compute the probability for an upwards fluctuation of the 
background to the measured test statistics.

If a definite signal prediction has to be checked, the value $R=r/2$ is
the appropriate choice.

For the determination of upper bounds the expected limit $n_{CL}$ 
can be minimized. An example is given in sect. \ref{appli2}.
This strategy works, if the background is sufficiently large. 

\subsection{Two discriminating variables}
\label{w2}
In the case of two weakly correlated physical variables like
particle mass $m$ and likelihood $L$, a likelihood inspired
definition of $\xi$, following equation (\ref{wdef}), is
\begin{equation}
\label{x2def}
\xi = \frac{D_{sm}(m) D_{sL}(L)}{D_{sm}(m) D_{sL}(L) + 2 D_{bm}(m) D_{bL}(L)}
\end{equation}
where the $D$'s are probility density functions and the indices
indicate the physical observables and signal (s) or background (b).
This procedure has been used in Higgs searches of the OPAL 
collaboration \cite{opal2}. 
Equivalently, the $\xi$ definition may be based onto formula (\ref{wrdef}).
If a larger correlation between $m$ and $L$ exists, it can be reduced 
by a linear transformation in the $m-L$ space before applying (\ref{x2def}).
  
The common use of eqs.(\ref{x2def}) and (\ref{wdef})
has the property that the weights $w_{ki}$ and the discriminating
variable $\xi_{ki}$ are identical, if the two physical variables are 
really uncorrelated, i.e. the product ansatz is correct. Any deviation 
indicates the presence of correlations or unacceptably large 
fluctuations in the Monte Carlo samples used to generate 
the histograms.
\footnote{Indeed the observation of a few anomalies triggered
additional Monte Carlo simulations in ref.\cite{opal2}}

\subsection{Related approaches}
An alternative approach, quite often used, is the ordering of
experiments according to the likelihood ratio $L_{sb}/L_b$ 
between the signal plus background and the background
interpretation of a data set (ref. \cite{read} to \cite{hu}).
Poisson statistics gives for this ratio
\begin{equation}
\label{LR}
L_{sb}/L_b = 
\exp (-r) \frac {\prod_{k,i} (s_{ki}+b_{ki})^{n(k,i)}}
                {\prod_{k,i} b_{ki}^{n(k,i)}} 
\end{equation}
where $r$  is the total signal rate and $n(k,i)$ is the number of 
candidates observed in the bin combination $(k,i)$.
The likelihood ratio method is equivalent 
to a weighted event counting with the filter function
\begin{equation}
\label{wML}
w_{ki} = \ln (1 + \frac{s_{ki}}{b_{ki}})
\end{equation}
The power expansion in terms of the signal to background ratio is
\begin{eqnarray}
w_{ki} = \frac{s_{ki}}{b_{ki}} - \frac{1}{2} \frac{s_{ki}^2}{b_{ki}^2}
+  \frac{1}{3} \frac{s_{ki}^3}{b_{ki}^3} +... \nonumber
\end{eqnarray}
This can be compared with twice the expansion of eq.(\ref{wdef})
It turns out that the first two terms agree and
the difference of the third terms is $s_{ki}^3/(12 \cdot b_{ki}^3)$  
only so that the results of both methods will be very similar in most
cases.

Significant differences are possible, if one or more candidates
are present in phase space regions where $s_{ki} >> b_{ki}$.
\footnote{An effect of this type, introduced by one candidate, is
visible in an earlier LEP combination of Higgs searches
\cite{LEP99}. It had no impact on the final result because the
candidate mass lies well above the combined mass limit.}.
 
Because eq.(\ref{wML}) has a singularity, it can produce spurious
discoveries, if the background distribution has a systematic
fluctuation in it, which is not handled properly in the statistical
analysis. The methods presented here do not check at all whether 
the underlying 
distributions $\epsilon_{ki},b_{ki}$ are consistent with the observed 
pattern. They take the theoretical distributions for shure
and ignore the fact that in a very low background region the 
systematic background error may be substantial. 
Contrary to (\ref{wML}) ,eq.(\ref{wrdef}) approaches a  
constant event weight in the limit $b_{ki} \rightarrow 0$
and is thus robust against this kind of effects.

It is an important advantage of (\ref{wrdef}) that it can 
be generalized to incorporate systematic errors, which destroy 
the statistical independency between $\xi$ bins, assumed in 
eq.(\ref{LR}).

Definition (\ref{wrdef}) is related to the maximum likelihood fit of the
signal rate. The logarithmic derivate of the likelihood is
\begin{eqnarray}
\frac{d \ln L_{sb}}{dr} = \frac{1}{L_{sb}} \cdot \frac{d L_{sb}}{dr} =
\frac{X}{r} -1
\nonumber
\end{eqnarray}
with
\begin{eqnarray}
  X =  \sum_{l} \frac{\epsilon_{k(l)} \cdot r} 
 {\epsilon_{k(l)} \cdot r + b_{k(l)}} \nonumber
\end{eqnarray}
which is equivalent to (\ref{wrdef}) with $R=r$.
The likelihood fit determines $r$ from the condition $X=r$.
  
\section{Weight distributions}
\subsection{Folding procedure}
\label{profol}
After the weight function $w(\xi)$ has been specified, 
density distributions $D(\xi)$ have to be transformed into 
distributions of $w$, called $P_1(w)$ for one event. The symbol $D$
stands for $s$ or $b$.
The histogram conversion is illustrated in fig.\ref{wconv}.
The cumulated integral $\int_0^{w_{cut}} P(w) dw $ 
at a special value $w_{cut}$ is illustrated by the shadowed
area. In case of histograms, all $\xi$ bins with 
$w_{ki} \le w_{cut}$ have to be counted.
A cumulated spectrum can be converted
into the differential one by taking bin-to-bin differences.
The central $w$ values of the bins will be assigned to all predicted
and observed events in that bin.
The analytic formula for a continuous function is
\begin{equation}
\label{xitow}
P_1(w) = \sum_l \frac{D(\xi_l)}{|\frac{d w}{d \xi }(\xi=\xi_l)|}
\end{equation}
The sum appears because the backward transformation from $w$ to
$\xi$ is not unique. All solutions $\xi_l$ of the equation $w(\xi)=w$
contribute.

Differential histograms $P_1(w_j)$ may have many gaps, 
but these are never populated by Monte Carlo or data events. The 
extremest case would be a delta function at $w=1$ for 
simple event counting. Because the distributions are not constant
inside the bins, binning effects can finally introduce relative 
errors of the order of 1/($<w>\cdot$number of bins)
in rate limits.

The distribution of $X= \sum_{l=1}^n w_{k(l)i(l)}$ for a fixed number of
$n$ events can now be computed from the distribution for one
event by iterative folding:
\begin{equation}
\label{folproc}
P_n(X) = \int_{{\rm{max}}(0,X-(n-1){\rm{max}}(w))}
^{{\rm{min}}(1,X)} 
P_{n-1}(X - w) \cdot P_1(w) \cdot dw
\end{equation}
The integration limits garantee that the arguments can not become
negative or exceed their upper limits. In general, 
these equations have no analytic solutions and must be evaluated 
numerically by matrix multiplication. The stepwise evolution of 
$P_n$ for a Gaussian signal and constant background is shown in 
fig. \ref{evol}. The singularity at the upper end of $P_1$, 
due to the maximum in the $\xi$ distribution, survives as a step 
for $n=2$ and as a vertical slope at the upper end for $n=3$. 
At $n=4$ all discontinuities have disappeared.

If the rates are large, many folding operations are necessary, but 
the results are needed for an $n$ interval only, whose lower and upper
bounds $n_{min},n_{max}$ have to be selected to reach
a requested accuracy.
It is then a faster procedure, to double the
event numbers in every folding step until the minimal value of $n$ is
reached, and to keep the distributions for
$n=2^m$ with integer $m$ for subsequent use. 
It is not necessary to compute folding integrals 
for any $n$. Distributions in the high 
$n$ region can be computed partly with interpolations because the shapes
are almost stable. To speed up numerical computations, it is also 
possible to combine two histogram bins into one, if the number 
of $X$ bins per standard deviation exceeds a cut with increasing $n$.
This process can be iterated.

Finally the Poisson distribution for appearance of $n$ events
has to be taken into account.
If $\overline{n}$ is the mean rate, the
final probability density is
\begin{equation}
\label{Pfol}
P(X) = \exp (-\overline{n}) \cdot \delta(X) +
 \sum_{n \ge X/{\rm{max}}(w)}^{\infty} \exp (-\overline{n}) \cdot
\frac{\overline{n}^n}{n!} \cdot P_n(X)
\end{equation}
For given $X>0$, only the terms with $n\ge X/$max$(w)$ can contribute.

The last formula is used to compute the complete distribution 
function $P_b(X)$ for background events.

It can be written down for signal events too, and the result
$P_s(X)$ has to be folded with $P_b(X)$ to get the 
overall distribution for signal and background, $P_{sb}(X)$.

It would be a nasty job to repeat many folding operations, 
if the signal rate $r$ has to be modified iteratively. 
Therefore, the $P_n$ distributions for fixed numbers of signal events, 
now called $P_{sn}$, are folded with the complete background 
distribution $P_b(X)$ from (\ref{Pfol}). To compute 
confidence levels, only the cumulated distributions are needed:
\begin{eqnarray}
\label{bkgadd}
C_n(X) = \int_0^{X} dZ \cdot
\int_0^{{\rm{min}}(Z,n \cdot {\rm{max}}(w))} dY 
            \cdot P_b(Z-Y) P_{sn}(Y)
\end{eqnarray}
The cumulated distribution for the sum of signal and
background is then
\begin{equation} 
\label{clsb}
CL_{sb}(X) = \int_0^X P_{sb}(Y) d Y 
=  \exp (-r) \cdot ( \exp(-\sum_{ki} b_{ki}) + \sum_{n=n_{min}}^{n=n_{max}} 
         \frac{r^n}{n!} \cdot C_n(X) )
\end{equation}
These results can now be used to compute the expected confidence
levels (\ref{aclsb}),(\ref{eclb}) and (\ref{eclsb}), 
needed to tune the $R$ parameter:
\begin{eqnarray}
E[CL_{sb}]_b = CL_{sb}(X_{cut}) 
     \hspace*{1cm} {\rm{with}} \hspace*{0.3cm}CL_b(X_{cut}) = CL 
     \nonumber \\
E[CL_b]_{sb} = CL_b(X_{cut})
     \hspace*{1.0cm} {\rm{with}} \hspace*{0.2cm}CL_{sb}(X_{cut}) = CL
     \nonumber
\end{eqnarray}
The parameter $CL$ is the request in criterion (iii) or (iv)
and $X_{cut}$ has to be computed from it by inversion of (\ref{CLdef}).

The expectation values needed for criteria (i) and (ii) are
\begin{eqnarray}
<CL_{sb}>_b = \int_0^{\infty} CL_{sb}(X) P_b(X) dX \nonumber \\  
<CL_b>_{sb} = \int_0^{\infty} CL_b(X) P_{sb}(X) dX \nonumber 
\end{eqnarray}
As already shown, both expectation values have their optimum
at the same value of $R$, which may now be a bit different 
from $r/2$.

An alternative method to compute the series of folding integrals
(\ref{clsb}) is given in ref.\cite{hu}, where fourier transformation 
is applied.

\subsection{Some analytic results}
The functions $P_1(w)$ and their statistical moments are known 
analytically in a few cases. All refer to the limit 
$R=0$, which means either a small signal to background 
ratio or the lowest probability for a background
fluctuation up to the median signal plus background level
(criterion (iii)). In any case a constant background is assumed.
\begin{itemize}
\item Gaussian signal. \newline
The $w$ distribution, its mean value and its mean square
for one signal event are
\begin{eqnarray}
\label{gauss1}
P_{s1}(w) = \frac{1}{\sqrt{ -\pi \cdot \ln w}} \hspace*{5mm}
<w>_s = \frac{1}{\sqrt{2}} \hspace*{5mm}
<w^2>_s = \frac{1}{\sqrt{3}}
\end{eqnarray}
At the signal maximum the weight is set to 1.
The background events are distributed according to
\begin{equation}
\label{gausb1}
P_{b1}(w) = {\mathcal{N}} \cdot \frac{\sqrt{2} \sigma_{\xi} \frac{d B}{d \xi}}
{w \cdot \sqrt{- \ln w}}
\end{equation}
This equation contains a normalization factor ${\mathcal{N}}$ and with
 ${\mathcal{N}}=1$ it gives the total background rate per $w$
 interval.
The width $\sigma_{\xi}$ refers to the signal and $dB/d \xi$ 
is the differential background rate.
The expression is not integrable at $w=0$, because an 
infinite number of events is taken into account far away from 
the signal. After truncation of the $\xi$ spectrum the integral converges.
The total mean and variance of $w$ are finite even without the
cutoff.  
\item Breit Wigner resonance over a constant background. \newline
The convention here is
\begin{eqnarray}
D(\xi) \sim \frac{1}{(\xi - \xi_0)^2 + \gamma^2} \nonumber
\end{eqnarray}
The distribution, the mean and mean square of $w$ for one signal
event are
\begin{eqnarray}
P_{s1}(w) = \frac{1}{\pi \cdot \sqrt{w \cdot (1 - w)}} \hspace*{5mm}
<w>_s = \frac{1}{2} \hspace*{5mm}
<w^2>_s = \frac{3}{8} \nonumber
\end{eqnarray}
The background distribution is
\begin{eqnarray}
P_{b1}(w) = {\mathcal{N}} \cdot 
\frac{\gamma \frac{d B}{d \xi}}{w \cdot \sqrt{w \cdot (1 - w)}}
\nonumber
\end{eqnarray}
\item Two-dimensional Gaussian. \newline
Two independent discriminating variables are distributed according
to \newline $D(\xi,\eta) \sim \exp (-(\xi - \xi_0)^2/
(2\sigma_{\xi}^2)) \cdot \exp (-(\eta - \eta_0)^2/(2 \sigma_{\eta}^2))$.
Instead of \newline eqs.(\ref{gauss1},\ref{gausb1}) one has
\begin{eqnarray} 
P_{s1}(w) = 1 \hspace*{5mm}
<w>_s = \frac{1}{2} \hspace*{5mm}
<w^2>_s = \frac{1}{3} 
\nonumber \\
P_{b1}(w) = {\mathcal{N}} \cdot \frac{2 \pi \sigma_{\xi} 
\sigma_{\eta}} {w} \cdot \frac{\partial^2 B}{\partial \xi \partial \eta}
\nonumber
\end{eqnarray}
\end{itemize}
From these results one gets the parameters needed for the high rate
estimates of confidence levels in sect.[2]:
\begin{itemize}
\item Gaussian signal
\begin{eqnarray}
 <X>_s = \frac{r}{\sqrt{2}} \hspace*{5mm} 
 <X>_b = \sqrt{2 \pi} \sigma_{\xi} \frac{d B}{d \xi}
 \hspace*{5mm} \sigma_s^2 = \frac{r}{\sqrt{3}} \hspace*{5mm}
 \sigma_b^2 = \sqrt{\pi} \sigma_{\xi} \frac{d B}{d \xi}
 \nonumber
\end{eqnarray}
\item Breit Wigner signal
\begin{eqnarray}
 <X>_s = \frac{r}{2} \hspace*{5mm} 
 <X>_b = \pi \gamma \frac{d B}{d \xi}  
 \hspace*{5mm} \sigma_s^2 = \frac{3}{8}r \hspace*{5mm}
 \sigma_b^2 = \frac{1}{2} \pi \gamma \frac{d B}{d \xi}
 \nonumber
\end{eqnarray}
\item Two-dimensional Gaussian
\begin{eqnarray}
 <X>_s = \frac{r}{2} \hspace*{5mm} 
 <X>_b = 2 \pi \sigma_{\xi} \sigma_{\eta} \frac{\partial^2 B}
{\partial \xi \partial \eta}
 \hspace*{5mm} \sigma_s^2 = \frac{r}{3} \hspace*{5mm}
 \sigma_b^2 = \pi \sigma_{\xi} \sigma_{\eta} \frac{\partial^2 B}
{\partial \xi \partial \eta}
 \nonumber
\end{eqnarray}
\end{itemize}

\section{Applications}
\subsection{Upper limits without background subtraction}
If nothing is known about size and spectral shape of the background,
upper limits for a signal rate can be obtained by 
ignoring the background in (\ref{clsb}).
The function $CL_{sb}$ has to be replaced by
\begin{equation} 
\label{clswob}
CL_s(X) = \int_0^X P_s(Y) d Y 
=  \exp (-r) \cdot (1 + \sum_{n=n_{min}}^{n=n_{max}} 
         \frac{r^n}{n!} \cdot
         \int_0^{{\rm{min}}(X,n \cdot {\rm{max}}(w))} dY \cdot 
         P_{sn}(Y) )
\end{equation}
Apart from trivial event counting the only meaningful ansatz 
for the weigths, valid for one search channel, is
\begin{equation} 
\label{stomax}
w_{ki} = \frac{s_{ki}}{{\rm{max}}(s_{ki})}
\end{equation}

The upper rate limit is obtained by solving (\ref{clswob}) for $r$,
if $CL_s$ is given. 

The 95\% exclusion limits ($CL_s=0.05$) for Gaussian and 
Breit-Wigner \newline $\xi$ distributions are shown as functions of the
test statistics in fig.\ref{n95X}. 
For comparison, the figure contains some dots marking the 95\% confidence
limits from Poisson statistics without spectral sensitivity.
In this case, the abscissa values are the observed event numbers.  

Fig. \ref{n95lim} shows a 95\% signal exclusion plot, which has been
computed from three observed events, using their measured masses
and varying a hypothetical reconance mass.
Accidentically, two of the mass values are almost identical.
The mass resolution is assumed to be Gaussian. 

The results obtained with eq.(\ref{clswob}) are given by  the solid
line. 
The curve has kinks at the rate limit 5.2.
This effect is visible in fig.\ref{n95X} too. It is due to the
singularity of the distribution $P_1(w)$ at $w=1$ (see fig. \ref{evol}). 
At the positions of the candidates, the rate limits
are slightly worse compared to the Poissonian limits of 4.74 for 
one and 6.30 for two observed events.  
These more pessimistic results from fractional counting
arise from theoretical configurations with low test statistics, 
which have more events in it than the observed numbers one and two.
This is the prize one has to pay for mass selectivity. In mass regions
away from the observed candidates the rate limits from fractional 
counting are more stringent than the Poissonian limits.
 
The problem of getting mass selective rate limits without
background subtraction has been addressed earlier.
Gross and Yepes \cite{yepes} use fractional event counting too. 
The weight is defined as the probability that an arbitrary event
has a larger mass difference with respect to the hypothetical 
particle than the candidate.
In the original publication the incorrect assumption had been made 
that the confidence limit for an integer number of fractional counts 
is equal to the Poissonian limit. The exclusions were too stringent.
Nevertheless, the ansatz for the weight is a legal alternative, 
and the rate limits obtained with it, 
using the folding procedure (\ref{clswob}), are added in
fig. \ref{n95lim}. The algorithm produces sharp spikes at the
candidate masses.

Another formalism to construct confidence levels 
was given by Grivaz and Diberder \cite{grivaz}.
They use a formula which looks
like the integral (\ref{clswob}), truncated at the number
of observed events:
\begin{eqnarray} 
E(n_{obs}) = \sum_{n=0}^{n_{obs}} \exp(-r) \cdot 
                      \frac{r^n}{n!} \cdot B_n \nonumber
\end{eqnarray}
Here, the $B_n$'s are probabilities that an arbitrary mass configuration
is less likely than the configuration of the $n$ measured events
closest to the hypothetical mass. The $B_n$'s are taken from the 
$\chi^2$ distributions of $n$ masses. The algorithm is not equivalent
to independent event counting. The authors have shown
that the expression can not be interpreted directly as a confidence
limit. It has some bias, which can be corrected for.
Numerical results are included in fig.\ref{n95lim}.
They are very similar to those of this work.

\subsection{Upper limits with background subtraction}
\label{appli2}

If the background is known without any systematic error, a
rate limit corresponding to a confidencve level $CL$ could be
determined from the condition
\begin{equation}
\label{rwsb}
CL = CL_{sb}(W_{tot}) 
\end{equation}
The $r$ dependence is given by (\ref{clsb}), which contains
the Poisson distribution.

As is well known, this procedure becomes problematic, if the
observed weight sum $W_{obs}$ is less than the expectation
from background. Equation (\ref{rwsb}) may have no
solution, which means that the frequency of appearance of the 
observation is less than $CL$ for any signal rate $r>0$. 
Mathematically, this is allowed and could be due to a statistical 
fluctuation. The problem may even survive if systematic errors
are added.

At this point a subjective element is introduced:
To garantee that even in exceptional cases the rate limit is 
conservative, the criterion for its determination is often sharpened 
to \cite{PDG,read,junk}:
\begin{itemize} 
\item The probability to observe a weight sum $X$ less than or equal
to the measured value $W_{obs}$, if the background contribution
alone is already $\le W_{obs}$, has to be less than $CL$.
\end{itemize}
This ansatz is motivated by the Bayesain treatment of background 
subtraction in counting experiments \cite{PDG} and it gives an 
overcoverage by definition.
To apply this condition, the following equation has to be solved for $r$:
\begin{eqnarray} 
CL = CL_s(W_{tot}) = \frac{CL_{sb}(W_{tot})}{CL_b(W_{tot})}
\label{rws}
\end{eqnarray}
Contrary to condition (\ref{rwsb}), this equation has a unique
solution for any value of $CL$.

Alternative prodedures have been published which avoid the
overcoverage as much as possible. The unified approach of Cousins 
and Feldman gives confidence belts instead of one-sided limits and has
been applied to the Poisson and the Gaussian distribution \cite{feldman}.
At low rates $r$, the confidence intervals are not central, and the 
upper limits are higher than those computed with ({\ref{rwsb}). 
The results are more stringent than those of (\ref{rws}).
Algorithms with optimized coverage for the Bayesian procedure have 
been investigated by Roe and Woodroofe and a connection to
(\ref{rws}) in the Poisson case has been found ~\cite{roe}.

The reason for adopting (\ref{rws}) is safetiness of upper limits.
If no event is observed at all, eq.(\ref{rws}) simplifies to 
$CL = \exp (-r)$, and any systematic background error cancels out.  

Figure \ref{n95r0scan} shows the 95\% exclusion limits on $r$ ($CL=0.05$)
for a Gaussian signal and constant background. The background level
is varied. It is parametrized by its mean contribution to the 
test statistics $\beta=<X>_b=\sqrt{2 \pi} \sigma_{\xi} \frac{d B}{d \xi}$.
Asymptotic limits can be obtained in the following way:
The expected background contribution $\beta$
can be subtracted from the observed test statistics $X$ before 
the rate limit is computed with \ref{clswob}. This would shift
the result without background subtraction by an amount 
$\frac{\beta}{<w>_s}=\sqrt{2}\beta$.
These asymptotic limits are reached for $ X > \beta$.
 
To get the results in fig.\ref{n95r0scan}, the weight definition 
(\ref{stomax}) was used again, which is equivalent to $R=0$ in 
eq.(\ref{wrdef}) 

According to the previous section, 
this is not the best choice for the filter. 
Figure \ref{ropt} shows the optimization of the
$R$ parameter for one special background level. It is based on the
median expected limit $E[n_{95}]_b$ , the rate $r$ which 
corresponds to $1-CL_s=0.95$, if background is observed at its median 
level $\beta$. 
Only a very weak dependence of $E[n_{95}]_b$ on $R$ of the order of
a few per mille can be seen. The limits from a real observation can 
vary by several \%, however, depending on the $\xi$ positions of the 
events, and in general they are a monotone function of $R$.

Fig. \ref{n95opt} gives the median expected limits as a function of
$\beta$. The lower curve indicates the $R$ parameters used to get 
these results. At large $\beta$, one has 
$R \approx 1/2 \cdot E[n_{95}]_b$. The difference to the above 
estimate $R \approx E[n_{95}]_b$ is due to the fact that finally
the limit computation is based onto $CL_s$ and not onto $CL_{sb}$.
Below $\beta=1$, the optimization becomes problematic.
Poisson fluctuations play a significant role.
There are several local minima of $E[n_{95}]_b$, if $R$ is varied,
and no solution has an obvious advantage over the other. 
The $R$ values given in fig.\ref{n95opt} have to be considered as
upper bounds on $R$; they are downward extrapolations consistent
with $R = 0.4 \cdot E[n_{95}]_b$, which is the unique result 
around $\beta=2$.

Fig. \ref{n95roptscan} is completely analog to fig.\ref{n95r0scan},
but now the optimized $R$ values from \newline fig. \ref{n95opt} are
used in the analysis. It should be noted that the test statistics $X$
is not the same in figs. \ref{n95r0scan} and \ref{n95roptscan}, and 
in the latter case it is also $\beta$ dependent. The dashed curve for 
$\beta=0$ corresponds to the Poisson distribution, because for any
finite $R$ and $\beta=0$ the algorithm does normal event counting.
 
For small finite $\beta$ the results depend strongly on $R$.
This ambiguity is illustrated in
fig. \ref{n95_3evtrvar}. The example of fig.\ref{n95lim}
with 3 measured particle masses is analysed again. This time it
is assumed that a background of 3 events is predicted within the
mass region of the plot, and this background is subtracted.
It corresponds to $\beta=0.88$. Three exclusion curves are shown,
which look completely different, but are all legal results.
The parameter $R=4$ gives an exclusion which looks somewhat
obscure, but it lies above the bound from fig.\ref{n95opt}, which
is approximately $R=1.6$. The second exclusion curve corresponds 
to this value. The third curve for $R=0$ corresponds, apart 
from background subtraction, to the result in fig. \ref{n95lim}.
This weight definition is known to be non-optimal.
In spite of this, it is recommended to keep the maximal mass 
resolution and to use $R=0$ in low statistics experiments.

\section{Confidence levels from the shapes of distributions}
The algorithms described in sect. [2] do not check the correctness of
the $\xi$ distributions. 
A large value of of a measured test statistics $X_{obs}$, normally
indicating a discovery, might also be due to an accumulation of 
mismodelled low weight background events. A statistical test which 
does not compare the total observed rate with a prediction and is 
sensitive to the local signal to background ratios  only, can be 
done in the following way:
The probability for an event, observed at $\xi_{ki}$, to be a signal 
event, is
\begin{eqnarray}
 p_{ki}=\frac{s_{ki}}{s_{ki}+b_{ki}} \nonumber
\end{eqnarray}
An arbitrary set of $n_{obs}$ events
obeys the polynomial distribution.
From the observed candidates a likelihood
\begin{eqnarray}
   L_{poly}=\prod_{l=1}^{n_{obs}} p_{k(l)i(l)} \nonumber
\end{eqnarray}
can be formed. A confidence level $CL_{poly}$ can be defined as the 
probability that an arbitrary experiment with the same number of candidates
gives at most the likelihood of the observed configuration.
This analysis can be done with
a background or a signal plus background interpretation of data.
Values of $CL_{poly}$ between 0.16 and 0.84 indicate consistency with
the tested model within 1 standard deviation. 
If $CL_{poly}$ has a normal value for the background interpretation,
but a low value for a signal plus background interpretation, 
a discovery is ruled out, even if $CL_b$ is close to 1. 
Vice versa, a large $CL_{poly}$ for the background interpretation 
supports a discovery.
If $CL_{poly}$ has normal values for both interpretations,
the test is not conclusive, either because the spectral shapes
of signal and background are similar or because the expected signal 
rate is too small.

To compute the confidence levels, the distribution functions of 
$L_{poly}$ are needed. The variable $L_{poly}$ can be replaced by its
logarithm. The test corresponds then to fractional counting of a fixed
number of events with
\begin{eqnarray}
w_{ki} = \ln \frac{s_{ki}}{s_{ki} + b_{ki}} \nonumber
\end{eqnarray}
The folding procedure is the same as in sect.[2].
The algorithm has the same disadvantage as the likelihood ratio
method: a singularity, this time at $s_{ki}$=0.
To avoid numerical problems, $p_{ki}$ has to exceed a minimum
value. A continuous upwards shift of this cut $p_{cut}$ removes
one candidate after the other from the sample, until the results 
are not anymore conclusive. The values of $CL_{poly}$ jump
at the discontiuities. 

As a toy example, fig.\ref{clpoly}a shows a Gaussian signal peak, a
linearly falling background and a pattern of candidates.
The mean values are 100 (background) and 20 (signal), 
the resolution is 15 bins. Compared to the background model, the
sample contains too many events (130).
The toy experiment is analysed at the hypothetical signal position.
The normal analysis gives confidence levels 
$CL_b=0.993$ and $CL_{sb}=0.45$, which might be a weak indication 
for a signal.
Fig.\ref{clpoly}c shows the confidence levels $CL_{poly}$ as 
function of $p_{cut}$. The smooth curves are two theoretical 
predictions: background events at the median level of the test
statistics
are analysed in terms of signal and background (lower curve) and
signal plus background is investigated assuming all events are
background (upper curve). The number of accepted events as a 
function of $p_{cut}$ is given too. The falling sensitivity with 
decreasing number of events is obvious from the pictures. Over all,
the comparison of the observed $CL_{poly}$ distributions with the 
median expectations shows somewhat better agreement with the 
background prediction and the test does not support a signal 
interpretation. Probably the background is underestimated.

\section{Systematic errors}
\subsection{Parametrization of systematic errors}
\label{esys}
The treatment here is limited to symmetric systematic errors,
described by Gaussian distributions. 
As a consequence, confidence levels shifts are proportional to
the mean squares of errors, if the latter are small.
Asymmetric errors modify the expectation values 
$<X>_b$ and $<X>_{sb}$ in first order and have larger 
impacts.

The errors are classified according to sources $j$. In principle 
every source may influence the $\xi$ spectra of signal and 
background in all channels.
It is parametrized by error functions
$\sigma^{(s)}_{j,ki}$ and $\sigma^{(b)}_{j,ki}$,  
whose absolute values are the rms errors, given binwise.

For the technical handling the following rules are introduced:
\begin{itemize}
\item Errors from the same source are treated as fully correlated 
between different bins of a signal or background histogram.
The signs of the error functions give the signs of the correlations.
\item Errors from the same source are treated as fully correlated 
between signal and background.
\item Errors from the same source are treated as fully correlated
between different search channels.
\item Errors from different sources are treated as completely 
uncorrelated.
\item The total relative error is much less than 100\%.
\end{itemize}
One comment on the independency of error sources is indicated.
It could be that the spectra $s_{ki}$ and $b_{ki}$
are available in an analytic form depending on parameters with 
correlated systematic errors. The error matrix can be diagonalized
to remove the correlations.
The assumption of independent sources is therefore no limitation.

Examples for complete independency are statistical uncertainties of 
Monte Carlo simulations. Considering signal and background, the 
number of error sources is twice the number of channels.

The last assumption on the error size is somewhat critical.
For instance, the error due to a mass resolution becomes
asymmetric far away from the mass peak and it has the same order
of magnitude as the spectrum itself. However, it will be shown later
that bins, where this happens, may be dropped anyway.

The effect of systematic errors on confidence levels are most 
easily studied with Monte Carlo simulations. To this aim, the input
spectra have to be modified according to
\begin{eqnarray}
\label{sysshift}
s^*_{ki} = s_{ki} + \sum_j \sigma^{(s)}_{j,ki} \zeta_j \\
b^*_{ki} = b_{ki} + \sum_j \sigma^{(b)}_{j,ki} \zeta_j
\nonumber
\end{eqnarray}
where the $\zeta_j$ are Gaussian random numbers of mean zero and
variance one. 

A major problem of eqs.(\ref{sysshift}) is the fact that error functions 
corresponding to likelihood or neural network variables are not 
well known, if known at all. Usually, systematic errors are evaluated
by modifying Monte Carlo simulations and counting the rate changes
above an effective selection cut. In this situation, an additional 
approximation is needed:
\begin{itemize}
\item For a given channel, the errors have the same dependence on the 
discriminating variable as the signal or background distributions:
\begin{eqnarray}
\label{esimpl}
\sigma^{(s)}_{j,ki} = \delta^{(s)}_{jk} s_{ki} \\
\sigma^{(b)}_{j,ki} = \delta^{(b)}_{jk} b_{ki} \nonumber
\end{eqnarray}
\end{itemize}
Here, relative errors $\delta^{(s)}_{jk}, \delta^{(b)}_{jk}$ 
are introduced which are source and channel specific, 
but bin independent. In general this ansatz is not true and 
dictated by lack of knowledge. Nevertheless it has been applied  
in searches (for the Higgs boson, see ref.\cite{LEP03} and
references therein).

\subsection{Correction of confidence levels in the frequentist
approach}
\label{clint}
In this subsection it is assumed that the test statistics $X$ is
a continuous variable. The distribution functions $P_{sb}(X)$ 
and $P_b(X)$ are not allowed to have delta function like singularities. 

In the following considerations, the event source and the analysis 
hypothesis are the same: background events are analysed 
in terms of background, the analogous is done for the combination 
of signal and background. The indices at the functions $CL_{sb},CL_b$
are dropped for simplicity.

If correct production rates are used and the analysis hypothesis is
correct, the $CL$ values are uniformly distributed between zero and 
one. This is not true if many potential observers use
parameters shifted by systematic errors. The distribution functions
$D(CL_{rec})$ of reconstructed confidence levels $CL_{rec}$ get 
observer dependent slopes. Because $CL_{rec}$ has a lower and an 
upper bound, the function $D_{rec}(CL_{rec})$, averaged over 
all observers, peaks at 0 and 1. 
Without any correction observers claim to often data deficits
or excesses, as illustrated in fig.\ref{gsys}.  

It is obvious how a correction can be done (see fig.\ref{gsys}): 
The distribution
$D_{rec}$ has to be integrated up to the reconstructed confidence
level $CL_{obs}$ of a certain observer and the integral is the corrected
confidence level:
\begin{equation}
\label{freqsys}
CL_{corr} = \int_0^{CL_{obs}} D_{rec}(CL_{rec}) d CL_{rec}
\end{equation}
This procedure has to be applied independently to $CL_b$ and $CL_{sb}$.
The problem is now that an individual observer can not reconstruct
the function $D_{rec}$, because it is based on smearing of the true
physical parameters, which are unknown. The observer 
will take his own spectra  $s(\xi)$, $b(\xi)$ instead of the 
true ones to evaluate the correction of $CL_{obs}$.
This replacement is unavoidable and causes deviations
of the average $CL_{corr}$ distribution from uniformity.

In the high rate approximation with Gaussian $X$ distributions,
the ansatz (\ref{freqsys}) should reproduce the naive result that 
statistical and systematic errors have to be added in 
quadrature. This is the case indeed, but the proof needs an 
additional assumption.

As already mentioned, an observer will start with original signal
and background spectra $s_{ki},b_{ki}$, from which the weights
$X_o$ are contructed. The mean value and the rms error of $X_o$ are
denoted by $<X_o>$ and $\sigma_o$, respectively. The signal and
background distributions will be modified according to 
eqs.(\ref{sysshift}). The new spectra are used to redefine the 
event weights. The original spectra can be inserted into 
eqs.(\ref{meandef}) with the modification that the new filter function
is used. The mean value of the test statistics $<X_o>$ and its rms
error  $\sigma_o$ will be shifted to $<X>$ and $\sigma$. 
The complete folding according to sect.\ref{profol} gives the function
$CL_{orig}(X)$, which expresses the original confidence levels
as a function of the modified test statistics $X$.
The modified spectra have to be analysed too, resulting in the 
the distributions $P_{rec}(X,\zeta)$ with the
statistical parameters $<X^*>$, $\sigma^*$ and its integral
$CL_{rec}(X)=\int_{-\infty}^{X} P_{rec}(Y,\zeta) \cdot dY $.
For clarity, it is written down explicitly that $P_{rec}$ depends on
the Monte Carlo variables $\zeta$.

The frequency distribution of $CL_{orig}$ is constant by definition,
because it describes the outcome of repeated measurements analysed 
with the same weight function. Any modified spectrum gives therefore 
a contribution 
\begin{eqnarray}
D_{rec}(CL_{rec}) d CL_{rec} =
d CL_{orig} = \frac{d CL_{orig}}{dX} \cdot dX \nonumber \\
\end{eqnarray}  
to the integral (\ref{freqsys}). 
The $CL_{rec}$ variable in the integration (\ref{freqsys}) may
by replaced by $X$, and a summation has to be performed over 
all Monte Carlo experiments. 
This leads to the final result
\begin{equation}
\label{sysgen}
CL_{corr}(CL_{rec}) = \int_{-\infty}^{\infty} d \zeta \cdot 
P_{sys}(\zeta) \cdot CL_{ori}(X^*)
\end{equation}
The integrand contains the parameter $X^*$. It has to be 
computed from the condition
\begin{equation}
\label{CLfix}
CL_{rec}=\int_{-\infty}^{X^*} P_{rec}(Y,\zeta) d Y
\end{equation}
which fixes the reconstructed confidence level to $CL_{rec}$.

One needs now a relationship between the shifted and original
distributions of the test statistics, $P_{rec}$ (for $\zeta \ne 0$) and 
$P_{orig}$ (at $\zeta=0$), and eq. (\ref{CLfix}) has to be solved.
Without loss of generality, we restrict $\zeta$ to one 
random variable.
Both distributions are approximated by Gaussians, and the ansatz for
its arguments, which is the mentioned assumption, is
\begin{eqnarray}
\frac{X-<X^*>}{\sigma^*} = \frac{X_o-<X_o>}{\sigma_o}
                  - \frac{\sigma_{sys}}{\sigma_o} \cdot \zeta
\nonumber
\end{eqnarray}
where $\sigma_{sys}$ is the systematic error of the expectation
value of the test statistics. The request of constant $CL_{rec}$,
eq.(\ref{CLfix}), leads to a linear relationship between the
integration limits $X^*$, $X_o$ and the random variable $\zeta$. 
Eq.(\ref{sysgen}) becomes
\begin{eqnarray}
CL_{corr} = \frac{1}{2 \pi} \int_{-\infty}^{\infty} d \zeta \cdot
\exp (-\zeta^2/2) \cdot
\int_{-\infty}^{X_o+\zeta \cdot \sigma_{sys}/\sigma_o} d Y
\exp(-\frac{(Y - <X_o>)^2}{2 \sigma_o^2})  
\nonumber
\end{eqnarray}
After a shift of the integration variable $Y$ and inversion of the
order of integrations one obtains the desired result
\begin{eqnarray}
CL_{corr} = \frac{1}{\sqrt{2 \pi (\sigma_o^2 + \sigma_{sys}^2) }}
\cdot \int_{-\infty}^{X_o} \exp (- \frac{-(Y - <X>)^2}
{2 (\sigma_o^2 + \sigma_{sys}^2)}) d Y \nonumber 
\end{eqnarray}

\subsection{Bayesian handling of systematic errors}
\label{bayhan}
Usually systematic errors are treated with the method
introduced by Cousins and Highland \cite{Cousins}. It is
trivial to generalize the Poissonian case, described in the original 
publications, to the situation with event discriminators in many 
channels. 

As mentioned, an observer does not know the true $\xi$ spectra, but only
his own estimates $s_{ki},b_{ki}$. The set $\zeta$ of stochastic
variables describes now the possible variants of the true spectra. 
Again a function $P_{sys}(\zeta)$ is introduced, which is here the 
Bayesian probability that the set $\zeta$ is the correct one.
The reconstructed confidence levels depend on $\zeta$.
Now two different statistical methods are mixed:
To include systematic errors,
the confidence levels from the frequentist approach are folded with
the observers believing about the true spectra:
\begin{equation}
\label{baysys}
CL_{corrected} = \int_{-\infty}^{+\infty} CL_{rec}(X,\zeta) 
\cdot P_{sys}(\zeta) d \zeta
\end{equation}
Here, $X$ is the measurement of the observer.

Theoretical spectra enter the analysis twice: they are needed to
construct the filter function and the absolute rates are used 
in the statistical analysis of sect.\ref{profol}. 
It is always a matter for debates whether systematic errors
should be assigned to the weight definition $w_{ki}$, and it 
became practice to keep this filter fixed \cite{LEP99,LEP03}. 
The argument for this is that the filter definition is arbitrary 
and, on statistical average, the results are correct for any fixed 
definition. Within the frequentist approach, this argument is not 
correct for a principle reason: It is logically impossible that 
different potential observers use the same numerical parameters for data
analysis. Every observer will construct his own filter function,
and the only agreement which could be reached, is a common value of
the ratio $R/r$ relevant for eq.(\ref{wrdef}). However, 
the comparison of the frequentist and the Bayesian approaches 
is simpler with a fixed weight definition, which is adopted in the 
following. One has therefore $X_o=X$ and $<X_o>=<X>$. 

Equations (\ref{baysys}) and (\ref{sysgen}) look completely different.
This rises the question, under which circumstances the Cousins
and Highland approach gives a uniform distribution of confidence levels.
 
A wide class of $X$ densities, for which both approaches agree,
can be constructed assuming shape invariance of the $X$ distribution:
\begin{equation}
\label{shape}
P_{rec}(X,\zeta) = F(\frac{X-<X^*>}{\sigma^*}) \hspace*{1 cm}
P_{orig}(X) = F(\frac{X-<X>}{\sigma}) 
\end{equation}
The denominators are the rms errors of the test statistics and $F$
is a common function. A constant reconstructed confidence level means
that the ratio \newline $(X-<X^*>)/\sigma^*$ is the same for 
$\zeta \ne 0$ and $\zeta=0$:
\begin{eqnarray}
\frac{X^*-<X^*>}{\sigma^*} = \frac{X-<X>}{\sigma} \nonumber \\
X^* = X \cdot \frac{\sigma^*}{\sigma} + <X^*> 
               - <X> \cdot \frac{\sigma^*}{\sigma} \label{xstar}
\end{eqnarray}
The ansatz for the systematic error within the frequentist approach is
\begin{equation}
\frac{<X^*>}{\sigma^*} = \frac{<X>}{\sigma} 
+ \zeta \cdot f(\zeta) \cdot \frac{\sigma_{sys}}{\sigma}
\label{ptrans}
\end{equation}
The $\zeta$ variable has a Gaussian distribution. The arbitrary 
function $f$ with \newline $f(0)=1$ describes non-Gaussian 
systematic errors. Equivalently, for the Bayesian treatment the
parametrization is
\begin{equation}
\frac{<X^*>}{\sigma^*} = \frac{<X>}{\sigma} 
+ \zeta \cdot g(\zeta) \cdot \frac{\sigma_{sys}}{\sigma}
\label{pptrans}
\end{equation}
A sufficient condition for the equivalence of 
(\ref{baysys}) and (\ref{sysgen}) is
\begin{equation}
P_{rec}(X,-\zeta) = P_{ori}(X^*)
\label{clsame}
\end{equation}
for any $X$ and $\zeta$. The minus sign on the left hand side appears
for the following reason:
The variable $\zeta$ parametrizes a shift from the original
distributions to the functions used by an arbitrary observer.
In the Bayesian interpretation, the direction of the shift has to be
inverted. 
Equations (\ref{shape}),(\ref{xstar}), (\ref{ptrans}) 
and (\ref{pptrans}) have now to be inserted into (\ref{clsame}).
Consistency is reached, if and only if $\sigma^*=\sigma$ and 
\newline $f(\zeta)=g(-\zeta)$. 
In general, this simple equivalence proof fails, if one
tries to introduce an $X$ dependence into the functions $f,g$ or to
violate the shape invariance (\ref{shape}).
The origin of the symmetry requirement on $f,g$ is illustrated in 
figure \ref{fig12}. 

The conclusion is that the equivalence between the 
frequenstist and the Bayesian treatments of systematic errors 
is partly very general, but there are also limitations:
\begin{itemize}
\item The distribution of the test statistics may be arbitrary.
\item The distribution of systematic shifts may be an arbitrary 
function.
\item However, equivalence is garanteed only, if systematic errors
shift the $X$ distributions, but keep their shapes, and
\item there is invariance of systematic shifts against
translations of the test statistics.
\end{itemize}
Apart from exceptions, the Bayesian treatment of systematic errors
does not agree with the frequentist approach, if one of 
the last two conditions is not fulfilled.  
Even if both approaches agree, the distribution of confidence levels,
averaged over many observers, are not nessessarily uniform, as
explained in the last subsection.

\subsection{Numerical treatment of systematic errors}

A repetition of the folding operations (\ref{folproc})
inside a Monte Carlo loop based onto (\ref{sysshift}) would be
a very time consuming analysis. It is a much simpler procedure to 
use the shape invariance (\ref{shape}) and the 
additivity assumption (\ref{ptrans}) with $g(\zeta)=1$.
 
The inclusion of systematic errors into the final results is then
straightforward: With the help of $N_{MC}$ Monte Carlo experiments
(\ref{sysshift}) and the definitions (\ref{meandef}),(\ref{vardef}) 
the systematic error is obtained from the mean 
$\chi^2$
\begin{eqnarray}
\chi_{sys}^2 = \frac{1}{N_{MC}} 
\sum_{{\rm MC \hspace*{1mm} experiments}}
(\frac{<X^*>}{\sigma^*}-\frac{<X>}{\sigma})^2 \nonumber
\end{eqnarray}
It has to be noted that this expression is written in a form
which garantees a cancellation of an arbitrary scaling factor in 
the $w_{ki}$, and that the expectation values involved can be
computed without the folding procedure (\ref{folproc}).

Equation (\ref{baysys}), together with (\ref{shape}) and 
\ref{ptrans}, leads to a folded distribution, from wich the corrected
confidence levels can be computed:
\begin{eqnarray}
P_{corr}(X) = \frac{1}{\sqrt{2 \pi}} \int_{-\infty}^{\infty} d \zeta 
\cdot \exp (-\zeta^2/2) \cdot P_{ori}
(X + \zeta \chi_{sys} \sigma) \nonumber \\
CL_{corr}(X) = \int_0^X dY \cdot P_{corr}(Y) \nonumber
\end{eqnarray}

The parameter $\chi_{sys}$ is different for background and a 
combination of signal and background and it depends 
on the overall signal to background ratio.
If $r$ has to be modified to find a rate limit, $\chi_{sys}$ has
to be reevaluated. 

The procedure has the advantage that it avoids a 
conceptual problem which exists otherwise for the extraction of rate 
limits from $CL_s$. Without systematic errors, $CL_{sb}$ is a monotone
function of $CL_b$, if the test statistics is eliminated. This
function becomes observer dependent in the presence of systematic
errors, which raises the question how $CL_s$ should be defined. 
In the above approach the ratio of folded functions 
$CL_{sb}$ and $CL_b$ is the natural choice. This method
has been suggested ealier for counting experiments by Zech \cite{zech}.

One has to keep in mind that the whole procedure is an approximate
one and can have biases.
   
\subsection{Poisson distribution at small rates}
The frequentist approach as introduced in sect.\ref{clint}
can not be applied to the Poisson distribution.
Here, the Bayesian ansatz has to be taken.
The Poisson distribution violates the criterion of shape
stability, as introduced in subsection \ref{bayhan}.

This raises the question whether the Bayesian
treatment gives a reasonable spectrum of reconstructed
confidence levels for the Poisson distribution at low rates.  
As an extreme case, which has nevertheless practical relevance
for background estimates, the problem has been studied for a 
very small mean rate $n_0=2$ with a big Gaussian systematic 
error of 20\%. The formalism how to get corrected confidence 
levels is described in ref.\cite{Cousins}.
For $n$ observed candidates the result is
\begin{eqnarray}
\nonumber
CL(n) = \sum_{i=0}^{i=n} I(i)
\end{eqnarray}
with
\begin{eqnarray} 
I(0)=\exp(-\overline{n_0} + \frac{1}{2} \sigma_{sys}^2)
\nonumber \\
I(1)= (\overline{n_0} - \sigma_{sys}^2) \cdot I(0) \nonumber \\
I(n)=\frac{\overline{n_0}-\sigma_{sys}^2}{n} \cdot I(n-1)
+\frac{\sigma_{sys}^2}{n} \cdot I(n-2) \nonumber
\end{eqnarray}
The following test has been done:
A set of potential observers was introduced with different 
assumptions on the mean rate. For any observer a new
Poisson distribution was generated and for any number of 
counts $n$ the corrected confidence levels $CL(n)$ were computed.
The entries in the overall $CL$ histogram were weighted with the
true probability to find $n$ counts.

Fig.\ref{pois} shows the distributions of confidence levels for
the special example. The differential spectrum of corrected 
confidence levels is not uniform at high $CL$, it has still a 
spike at $CL=1$. The right column shows the cumulative $CL$ 
distribution in a two-sided logarithmic representation. 
The result does not approach the diagonal at $CL=1$.
One could argue that the same relative error was assumed for all
potential observers and that a more adequate choice would be 
$\delta \sim 1/\sqrt{n_0}$. Tests have shown that this
ansatz gives little improvement but does not cure the problem.   

An exceptional case was recently published by Bityukov who
investigated statistical errors of Monte Carlo simulations as
source of systematic errors in counting experiments \cite{bit}. 
If the mean rate $n_0$ is taken from a simulation based onto 
Poisson statistics which has the same mean value as the data 
sample, the effect is absent. The criteria of shape stability
of the distribution and translational invariance of systematic
errors are violated and these effects cancel.

It is the conclusion that indications for discoveries
obtained from low statistics samples should be considered with care, 
if the background has a substantial uncertainty. Even after 
correction for systematic errors the significance of the 
observation is still overestimated and this bias has to be studied. 

\section{Event weighting with systematic errors}

In the preceding section, systematic errors have been added to the
final results, but the filter function (\ref{wrdef}) was 
optimized with respect to statistical errors only. 

If search channels with much different systematic errors are
combined or an uncertain amount of low weight background 
events contributes to fractional counting, this is not the best 
way to analyse data: bins with large systematic errors 
have to be downgraded. 

The procedure described in sect.\ref{evw} can be generalized to
do this. Again the limiting case of Gaussian distributions 
for the test statistics is considered. The generalization is 
straightforward for the three cases $R \rightarrow 0$, $R = r/2$ 
and $R=r$, which cover the range of $R$ values in formula (\ref{wrdef}).

\begin{itemize}
\item ($\alpha$) $R=r/2$. \newline
The optimization criteria (i),(ii) had the following form:
The probability that an arbitrary measurement of signal and
background events gives a total weight $X$ less than or equal to 
the weight of an arbitrary background sample, should have a 
minimum. This condition needs  now the supplement:
The total weights $X$ for the comparison are measured by 
independent, arbitrary observers.
\item  ($\beta$) $R=r$. \newline
This criterion minimized the fluctuation of signal and
backgrond down to the background expectation.
It had the form $<X>_s/\sigma_{sb}$=max.,
The systematic errors have to be included into the denominator now.
\item  ($\gamma$) $R=0$.
This criterion minimized the background fluctuation
up the signal plus background expectation.
The ratio  $<X>_s/\sigma_b$ has to be a maximum, with the
systematic errors included in $\sigma_b$.
\end{itemize}

Criterion ($\alpha$) is a bit more complicated than the others and means
\begin{eqnarray}
\sigma^2  / (\sum_{ki} w_{ki} s_{ki} )^2 
 = (\sigma_{sb}^2 + \sigma_b^2)/
 (\sum_{ki} w_{ki} s_{ki} )^2 = {\rm{min.}}  \nonumber
\end{eqnarray}
The contributions of systematic errors to the variances have to be 
computed with (\ref{meandef}) and (\ref{sysshift}):
\begin{eqnarray} 
\sigma_{sb}^2 = \sum_{ki} w_{ki}^2 \cdot (s_{ki} + b_{ki})
+ \sum_j (\frac{\partial<X>_{sb}}{\partial \zeta_j})^2 \nonumber \\ 
\sigma_b^2 = \sum_{ki} w_{ki}^2 \cdot b_{ki}
+ \sum_j (\frac{\partial<X>_b}{\partial \zeta_j})^2 \nonumber \\
\sigma_{sb}^2 = 
\sum_{ki} w_{ki}^2 (s_{ki} + b_{ki} ) + \sum_j ( \sum_{ki} w_{ki}
  ( \sigma^{(s)}_{j,ki} + \sigma^{(b)}_{j,ki} ))^2 \nonumber \\
\sigma_b^2 =  \sum_{ki} w_{ki}^2 b_{ki} +
\sum_j ( \sum_{ki} w_{ki} \sigma^{(b)}_{j,ki} )^2 \nonumber
\end{eqnarray}

The optimization leads to the following equations:
\begin{eqnarray}
\label{swopt}
w_{ki} \cdot (s_{ki} k_1 + b_{ki} k_1 + b_{ki} k_2) 
+ \sum_{lm} w_{lm} \cdot 
 \sum_j (\sigma^{(s)}_{j,lm} + \sigma^{(b)}_{j,lm})
        (\sigma^{(s)}_{j,ki} + \sigma^{(b)}_{j,ki}) \cdot k_1 
 \nonumber \\
+ \sum_{lm} w_{lm} \cdot \sum_j \sigma^{(b)}_{j,lm} \sigma^{(b)}_{j,ki}
 \cdot k_2 = s_{ki} \hspace*{2cm} 
\end{eqnarray}
The cases $(\beta)$ and $(\gamma)$ are included too:
one has $k_1=k_2=1$ for condition $(\alpha)$,$k_2=0$ for 
condition $(\beta)$ and $k_1=0$ for $(\gamma)$.
The double sums correct the weights (\ref{wrdef})
for systematic errors, but they contain the final result so that
the system of linear equations (\ref{swopt}) has to be solved.  

Among the weights one may find negative values. Mathematically
there is nothing wrong with this: The algorithm tries to extract
information on background from signal tails and to extrapolate this
into the signal region to improve the accuracy. However, because the 
errors on the shapes of $\xi$ distributions are not well known and 
were even ignored in (\ref{esimpl}), the appearance of negative weights 
is completely unacceptable. To drop bins with low signal content, 
equation (\ref{swopt}) can be supplemented by the request that
all $w_{ki}$ should be positive or 0.

Together with this condition, (\ref{swopt}) has a unique solution.

Let $N$ be the total number of histogram bins.
The normalization condition $X_s=\sum_{ki} w_{ki} s_{ki}$=const. 
defines an $(N-1)$-dimensional hyperplane in the space of weights 
$w_{ki}$. The $N$ inequalities
$w_{ki}\ge 0$ define an $(N-1)$ hyperplanar object with $N$ corners within
this hyperplane, a so called simplex. The simplest examples are a 
connection line for $N=2$, a triangle for $N=3$ and a tetraedron 
for $N=4$. At the corners only one of the $w_{ki}$ is greater than 0.
The surface of the simplex consists of $N$ hyperplanar
objects of dimension $(N-2)$, which are simplices again. The simplest
examples are the end points of the connection line for $N=2$, the
sites of the triangle for $N=3$ and the surface
triangles of the tetraedron for $N=4$. These surface elements are 
characterized by one vanishing $w_{ki}$. Two of the $(N-2)$-dimensional
surface elements have one $(N-3)$-dimensional simplex in common. 
There are $N \cdot (N-1) /2$ of these objects, on which two weights 
vanish. This decomposition can be repeated until one reaches the 
corners. All curvature components on these substructures vanish. 

The condition $\sigma^2  / (\sum_{ki} w_{ki} s_{ki} )^2=p$
defines an $N$-dimensional hyperellipsoid, whose size depends on the
constant $p$. For sufficiently small values all points
of the simplex $w_{ki}\ge 0$ lie outside the hyperellipsoid.
Because both the error ellipsoid and the simplex are convex
and all curvature components of the ellipsoid are non-zero,
there exists exactly one value of $p$, for which 
the simplex becomes a tangential object of the hyperellipsoid. 
The coordinates of the tangential point are the weights.

The point computed with (\ref{wrdef}) lies in the interior of
the $N-1$-dimensional simplex. In general, the error ellisoid 
containing it will have a larger value of $p$ than this solution.
The ansatz (\ref{swopt}) leads to a better
discrimination between hypotheses (A) and (B) with the same
optimization criterion, even if less bins are used.

This is illustrated in fig.\ref{wsysopt}, which shows expected upper 
rate limits for a Gaussian signal arising from a constant background.
On the left hand side, the original weights (\ref{wdef}) are compared
with the result of (\ref{swopt}). It turns out that the region of
accepted events around the signal peak is rather narrow, if the
systematic errors are comparable to the statistical ones. The
acceptance window
depends on the background level $\beta$, which is again the number of 
events in the $\xi$ interval $\sqrt{2 \pi} \sigma_{\xi}$.
The expected rate limits with the filter (\ref{swopt}) (full lines)
are lower than the limits computed with (\ref{wrdef}) (dotted lines).
It is also evident from the figure that the ordering of curves is 
opposite for the same filters, if the systematic errors are 
not included in the statistical analysis.

Results of similar quality can be obtained 
with (\ref{wrdef}) together with a cut on $s_{ki}/b_{ki}$. 
This would have the consequence that another parameter has to be 
tuned. From a principle point of view,
it would be some irony if a cut would be introduced here,
because fractional counting was partly introduced to avoid 
hard cuts in event acceptance.

The application of this weighting method is meaningful, if systematic
errors, including their correlations, have the same the order of 
magnitude as the statistical errors or are even larger.
A relevant physical example is the flavor-independent search
for Higgs bosons \cite{flavblind}. Compared to more specific 
Higgs searches, the background is larger, but systematic 
uncertainties are very similar.
Absolute upper limits grow with the square root
of background, but the systematic error is proportional to it,
so that systematic errors become important.

\section{Summary}

The method of fractional event counting has been 
presented. The statistical analysis uses the frequentist 
approach. A very simple weight function with one free parameter
was derived and it is described how it can be ajusted to get 
optimal separation of a physical signal from background.
It turned out that there is no saticfactory optimization strategy
for very low statistics experiments and it is proposed to use
simply the signal-to-background ratio or the signal shape as weight.
 
Very simple formulas are given to compute expected and observed
confidence levels in the high rate limit, and for very simple examples
like a Gaussian or a Breit-Wigner signal over a constant background
analytic results are presented.

A statistical test is suggested as a supplement to normal 
statistical analyses which is is based on polynomial 
statistics. It is sensitive to the ratio of signal 
and background spectra, but does not use the observed absolute 
rate for model comparisons.

The frequentist and the Bayesian treatments of systematic errors 
are compared for a continuous test statistics, whose distribution
has no local delta function like spikes.
Both approaches agree, if systematic errors introduce shifts 
of the distribution of test statistics without modification of 
its shape, and if the systematic shifts are invariant against 
translation of the test statistics.
 
It has been shown that the Bayesian treatment of systematic 
errors in low statistics experiments is problematic: the 
results may be biased and this has to be studied.

Finally a method was introduced to reduce the impact
of systematic errors on confidence levels. It includes systematic
errors in the event weights and does an automatic bin dropping to
tolerate that detailed spectral shapes of systematic errors 
are often not well known.

\section*{Acknowledgements}
This work would not have been possible without many stimulating
discussions in the Higgs working group of the OPAL collaboration
and also in the LEP wide working group on Higgs searches.
Especially the author thanks U.Jost for carrying out careful 
tests during the initial phase of this work. The author is
pleased to acknowledge the Bundesministerium f\"ur Forschung und 
Technologie for the support given to the OPAL project.

\section*{Appendix: Comments on the comparison of two 
arbitrary hypotheses}

In the preceding sections, the physical hypotheses (A) and (B) 
differed by an excess of events in one of them everywhere. 
Often physical models are parameter dependent with 
locally different signs of cross section shifts.
A simple example is the comparison of two angular correlations, 
where the measurement is based on a fixed number of events.

In the following, it is summarized briefly, how the weighting has to
be modified to handle the more general case.

Let be $a_{ki}$ and $b_{ki}$ the local rates. The previous results 
are reproduced with $a_{ki} = b_{ki} + s_{ki}$.
The weight optimization can be repeated with the normalization
$\sum w_{ki} \cdot (a_{ki}-b_{ki})=$const., and the result is
\begin{eqnarray}
w_{ki} = \frac{{\mathcal{U}} \cdot (a_{ki} - b_{ki})}
               {U \cdot a_{ki} + (1-U) \cdot b_{ki}} \nonumber
\end{eqnarray}
with a free parameter $U$ which replaces R. Similarly, 
${\mathcal{U}}$ is an arbitrary renormalization factor.
To garantee a positive denominator,$U$ should be constrained to
$0 \le U \le 1$. The weights can now become negative, but they
have a lower and an upper bound.  The folding procedures to get
the distributions of the test statistics are the same, but $X$ lies
now in the interval $-\infty$ to $\infty$ and the lower integration
limit in the confidence level integrals has to be set to a 
sufficiently large negative number. 

The statistical test for a fixed number of events, based onto the 
polynomial distribution, can be modified to use the polynomial 
likelihood ratio of the two models as discriminator. The event 
weights become then
\begin{eqnarray}
w_{ki} = \ln \frac {a_{ki}}{b_{ki}}
\nonumber
\end{eqnarray}
This formula is symmetric and has
singularities for $a_{ki}=0$ and $b_{ki}=0$; lower and 
upper cuts on the ratios of local rates are needed.
An example for its application is the mentioned angular
distribution check.

Finally, the weighting with systematic errors leads to the linear
equations \newpage
\begin{eqnarray}
\label{gwopt}
w_{ki} \cdot (a_{ki} \cdot k_1 + b_{ki} \cdot k_2) 
+ \sum_{lm} w_{lm} \cdot 
 \sum_j \sigma^{(a)}_{j,lm} \sigma^{(a)}_{j,ki} \cdot k_1
  \nonumber \\
+ \sum_{lm} w_{lm} \cdot \sum_j \sigma^{(b)}_{j,lm} \sigma^{(b)}_{j,ki}
 \cdot k_2 = a_{ki} - b_{ki} \nonumber 
\end{eqnarray}
The numerical factors $k_1,k_2$ are defined as before and
depend on the hypothesis one likes to verify.
 
The requirement of positive weights is meaningless. 
It was introduced to circumvent bad knowledge of the spectral 
shapes of systematic errors in regions where the difference
between the models is small. Here, bins with $a_{ki} \approx b_{ki}$ 
are not significant, and they can be dropped with the request
that $w_{ki}$ must have the same sign as $a_{ki} - b_{ki}$.

\begin{figure}[htb]
\begin{center}
\mbox{
\epsfig{file=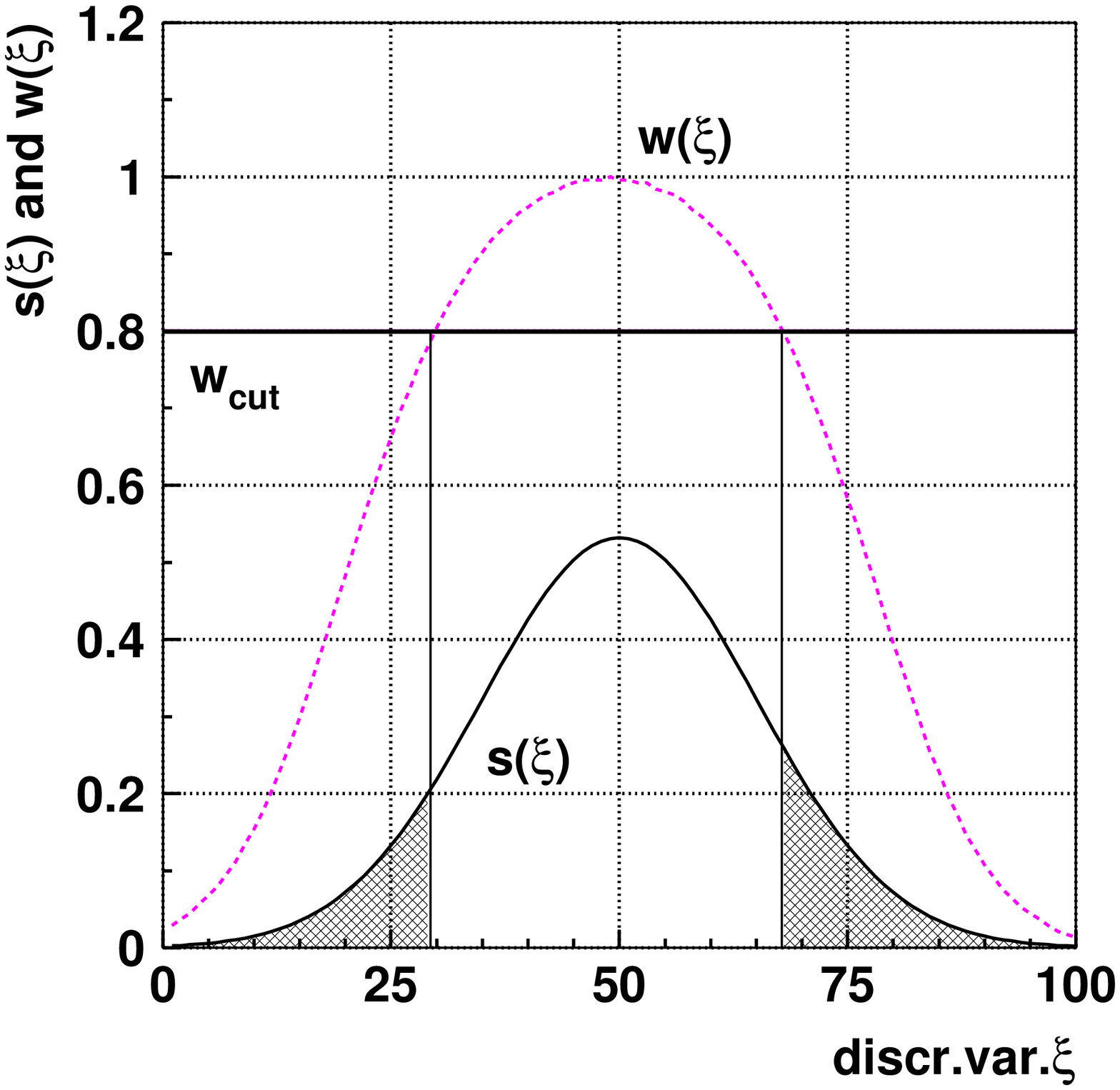,height=8cm}
     }
\caption {{\it Construction of the cumulated weight distribution
of signal events from their $\xi$ distribution and the weight function
$w(\xi)$.
}}
\label{wconv}

\mbox{
\epsfig{file=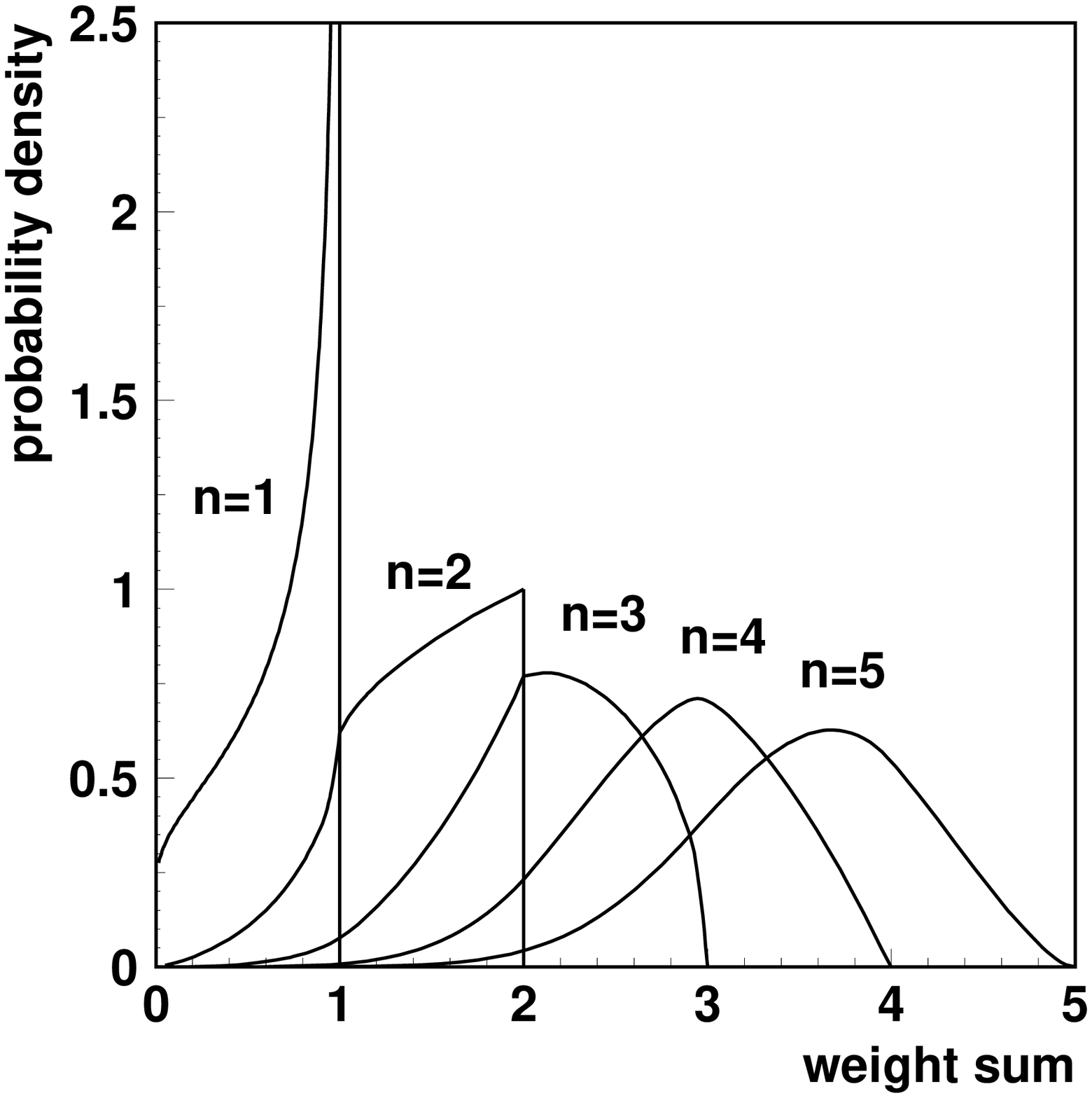,height=8cm}
     }
\caption {{\it Spectra of the test statistics $X$ for fixed numbers 
of events. The distributions are for small signal to background
ratio and a Gaussian signal over a constant background.The functions
are given for the signal.}}
\label{evol}
\end{center}
\end{figure}

\begin{figure}[htb] 
\begin{center}
\mbox{
\epsfig{file=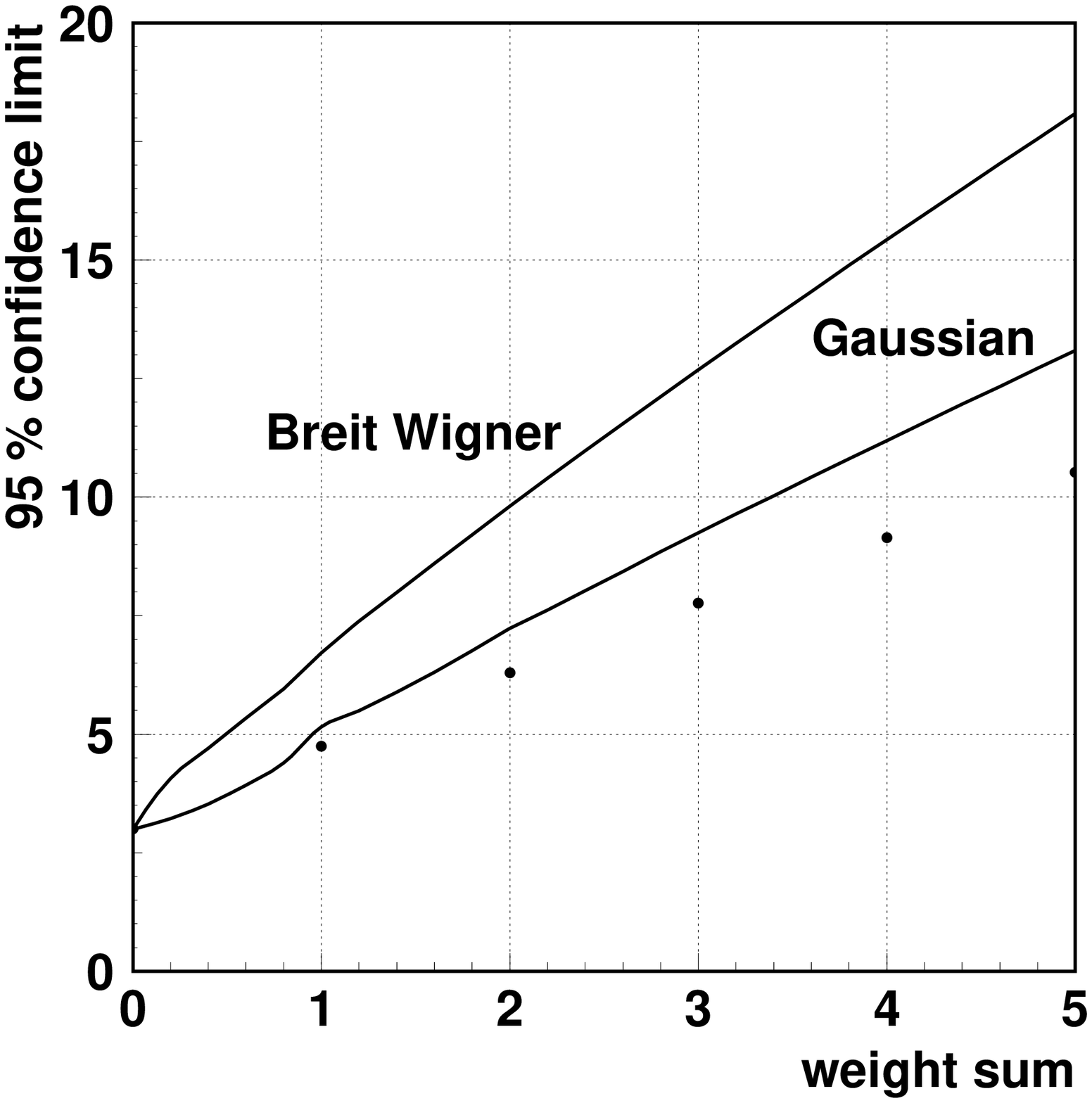,height=8cm}   
}
\caption{{\it Count rates excluded with 95\% confidence 
without background subtraction.
lower curve: Gaussian distribution, upper curve: Breit-Wigner 
resonance. The dots at integer abscissa values are the Poissonian 
limits from unweighted counting.}}
\label{n95X}

\mbox{
\epsfig{file=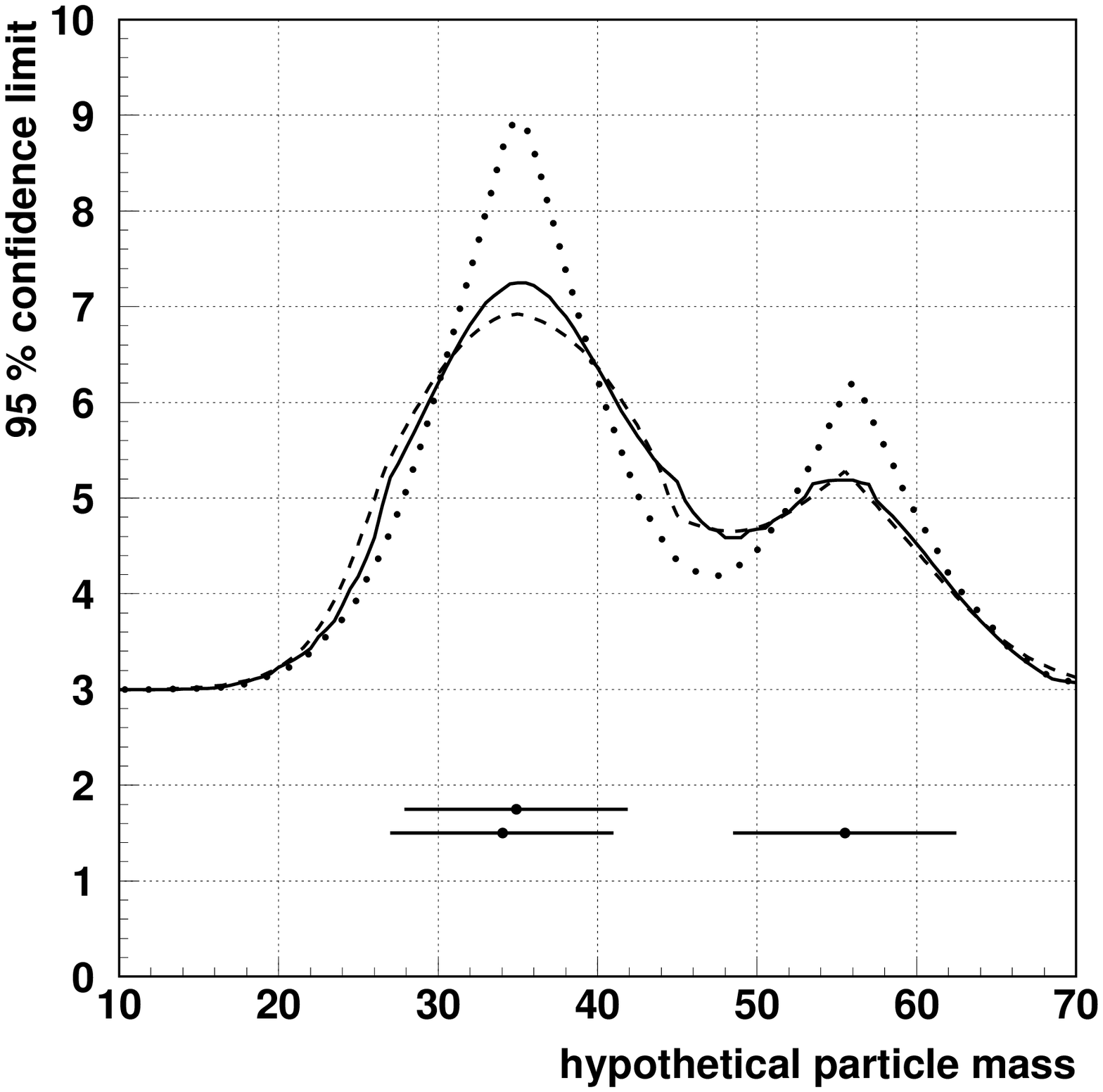,height=8cm}
     }
\caption {{\it Limits on signal production rates from 3 events without 
subtraction of background. A Gaussian mass spectrum is assumed.
The candidate positions are given by the points and the mass 
resolution is indicated by the error bars. Full curve: this work,
dashed curve: Grivaz and Diberder, dotted curve: weighting of 
Gross and Yepes.}}
\label{n95lim}
\end{center}
\end{figure}

\begin{figure}[htb] 
\begin{center}
\mbox{
\epsfig{file=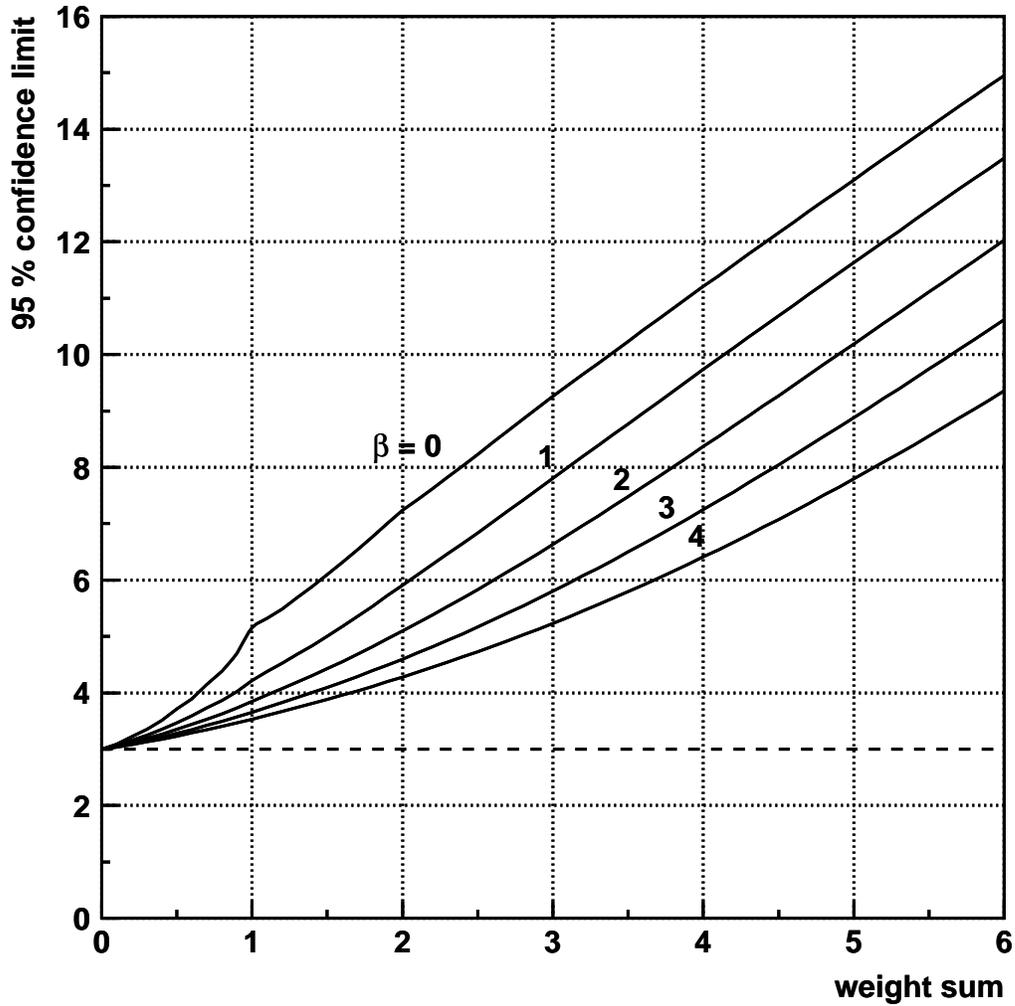,height=15cm}   
} 
\caption{{\it Count rates excluded with 95\% confidence 
as function of the weight sum. The background is subtracted.
The limits are for a Gaussian signal distribution and a constant 
background level $\beta$. The weight is taken proportional to the
signal to background ratio.}}
\label{n95r0scan}
\end{center}
\end{figure}   

\begin{figure}[htb] 
\begin{center}
\mbox{
\epsfig{file=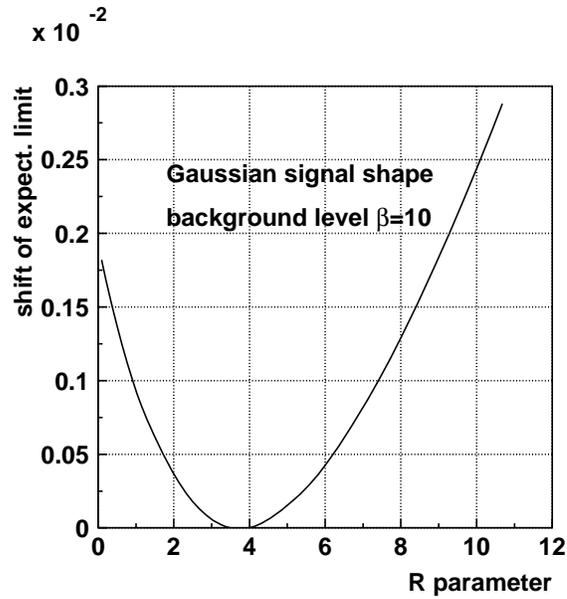,height=8cm}   
} 
\caption{{\it Dependence of the median expected 95\% confidence limit
on the rate parameter $R$. The background level is $\beta=10$.
}}
\label{ropt}
\end{center}
\end{figure}   

\begin{figure}[htb] 
\begin{center}
\mbox{
\epsfig{file=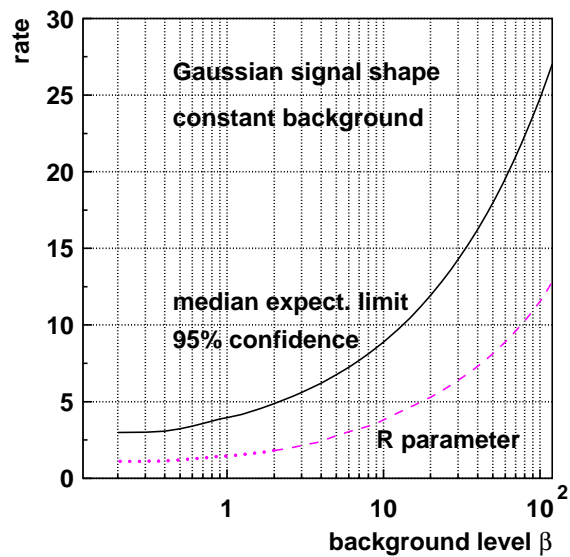,height=8cm}   
} \caption{{\it Median expected rate limits as a function of the 
background level $\beta$, if no signal exists. The lower curve gives 
the parameters $R$ used to compute the limits.
}}
\label{n95opt}
\end{center}
\end{figure}   

\begin{figure}[htb] 
\begin{center}
\mbox{
\epsfig{file=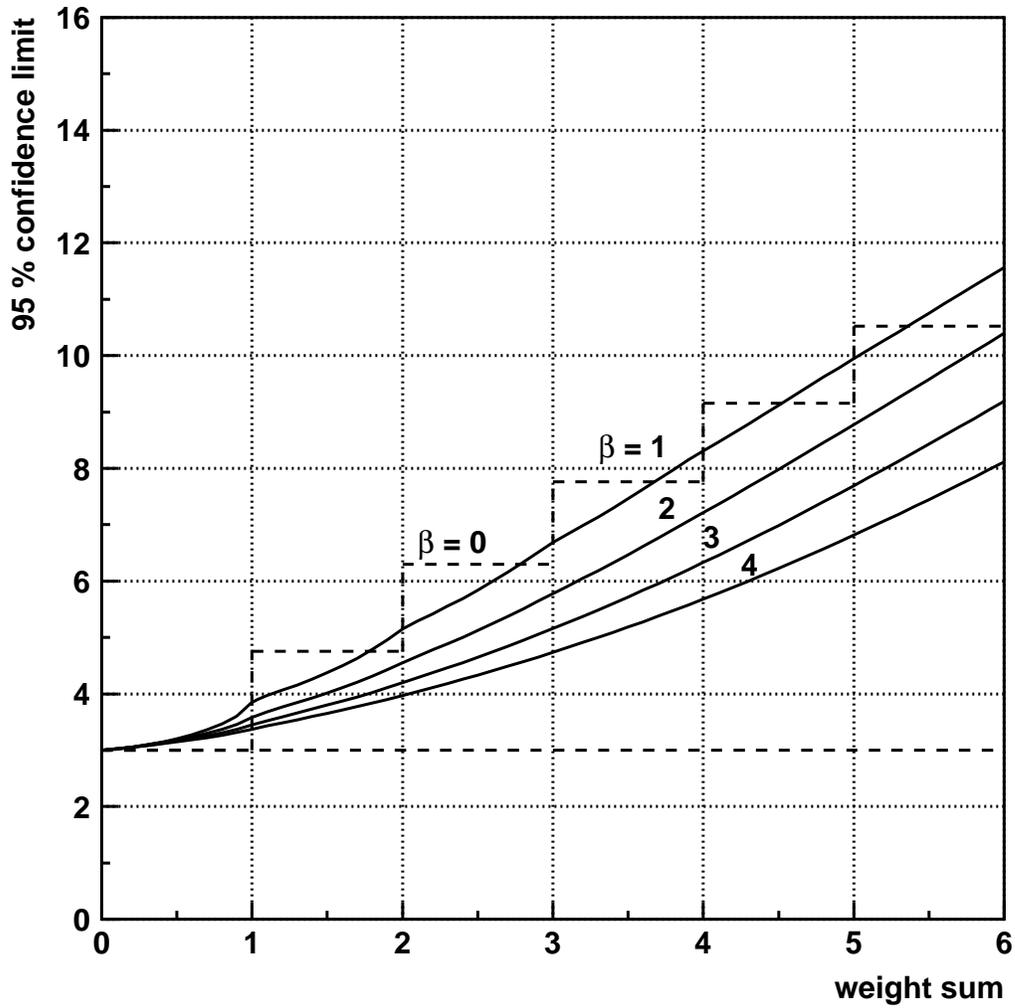,height=15cm}
}
\caption{{\it Count rates excluded with 95\% confidence 
as function of the weight sum. The background is subtracted.
The limits are for a Gaussian signal distribution and constant 
background levels $\beta$. The $R$ parameters of fig. \ref{ropt} were
used to define the weights (see text).
}}
\label{n95roptscan} 
\end{center}
\end{figure}   

\begin{figure}[htb]
\begin{center}
\mbox{
\epsfig{file=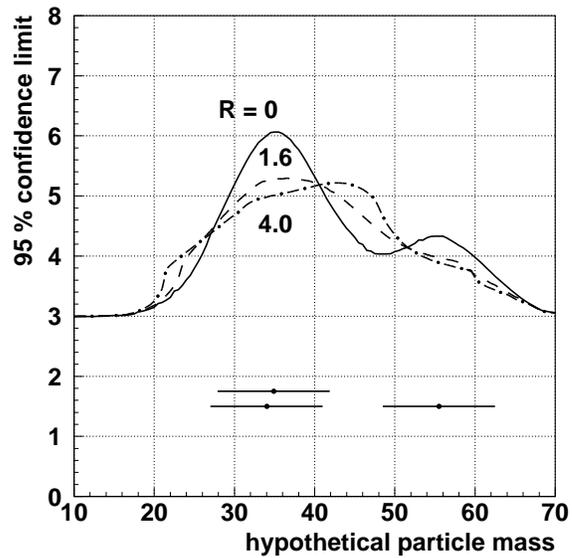,height=8cm}
     }
\caption {{\it Limits on the production rate from 3 observed events with
subtraction of 3 background events. The data are identical to 
fig. \ref{n95lim}. The curves are for different definitions of the
weight algorithm and demonstrate the ambiguities in the analysis. 
}}
\label{n95_3evtrvar}
\end{center}
\end{figure}

\begin{figure}[htb]
\begin{center}
\mbox{
\epsfig{file=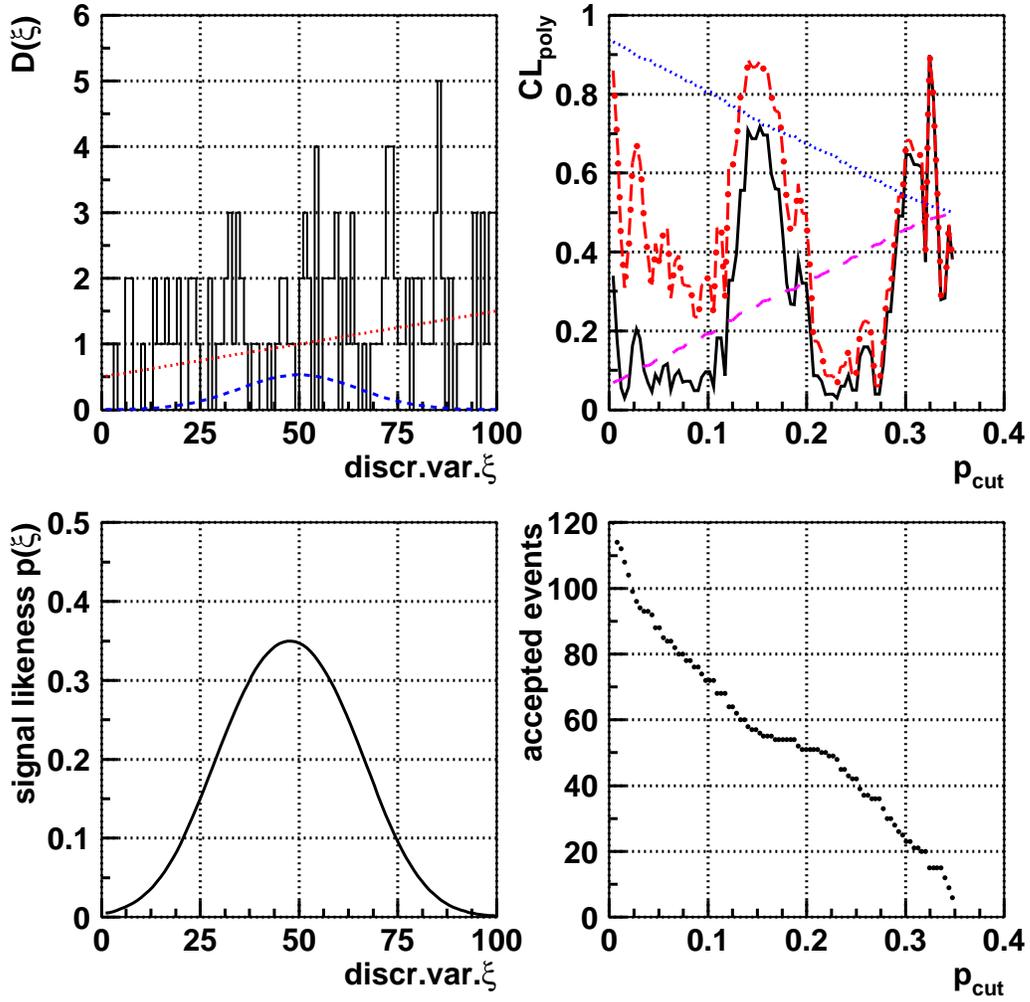,height=15cm}
     }
\caption{{\it Confidence levels based on the polynomial distribution
for a toy example. Upper left: Signal, background and
candidate distributions. Lower left: The signal probabilities $p(\xi)$.
Upper right: Confidence levels. The full and the dash-dotted curves
are for the 'data', the smooth curves are median expectations (see
text). The analysis assumptions are background for the upper curves,
signal and background for the lower curves. Lower right: 
Number of events with $p(\xi) \ge p_{cut}$. }}
\label{clpoly}
\end{center}
\end{figure}

\begin{figure}[htb]
\begin{center}
\mbox{
\epsfig{file=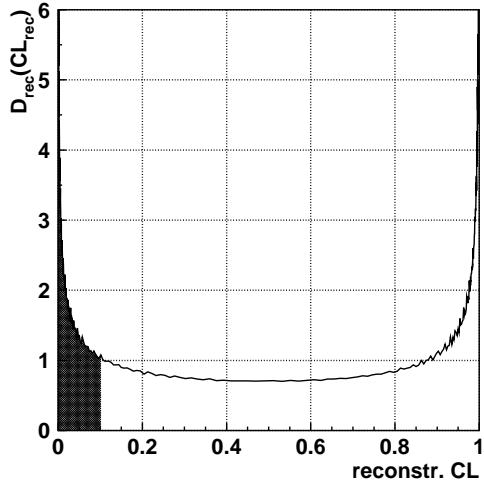,height=7cm}
     }
\caption {{\it Distribution of reconstructed confidence levels 
for a Poisson distribution with a mean rate of 100 events and 
a systematic error of 10\%, as reconstructed by many observers.
}}
\label{gsys}
\end{center}
\end{figure}

\begin{figure}[htb]
\begin{center}
\mbox{
\epsfig{file=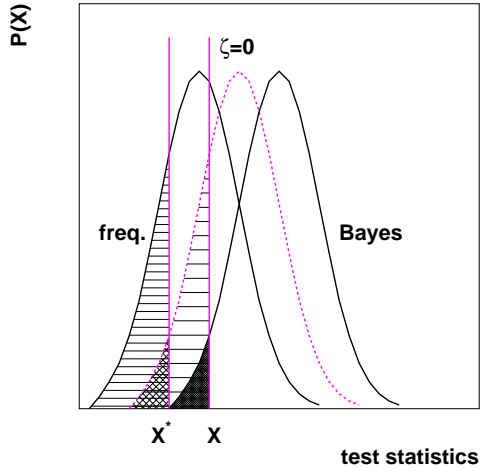,height=7cm}
     }
\caption {{\it Relationship between the frequentist and the Bayesian
treatment of systematic errors. Central curve: Original distribution
of the test statistics. $X$ is a measured value. 
Right curve: Shifted distribution according to Bayesian error
treatment for $\zeta<0$. Dark area: Contribution to the corrected 
confidence level. Left curve: Shifted distribution according to the
frequentist approach for the same value of $\zeta$. The two
horizontally hatched areas are equal by contruction.
The agreement of both approaches is garanteed if the small 
marked areas are equal .
}}
\label{fig12}
\end{center}
\end{figure}

\begin{figure}[htb]
\begin{center}
\mbox{
\epsfig{file=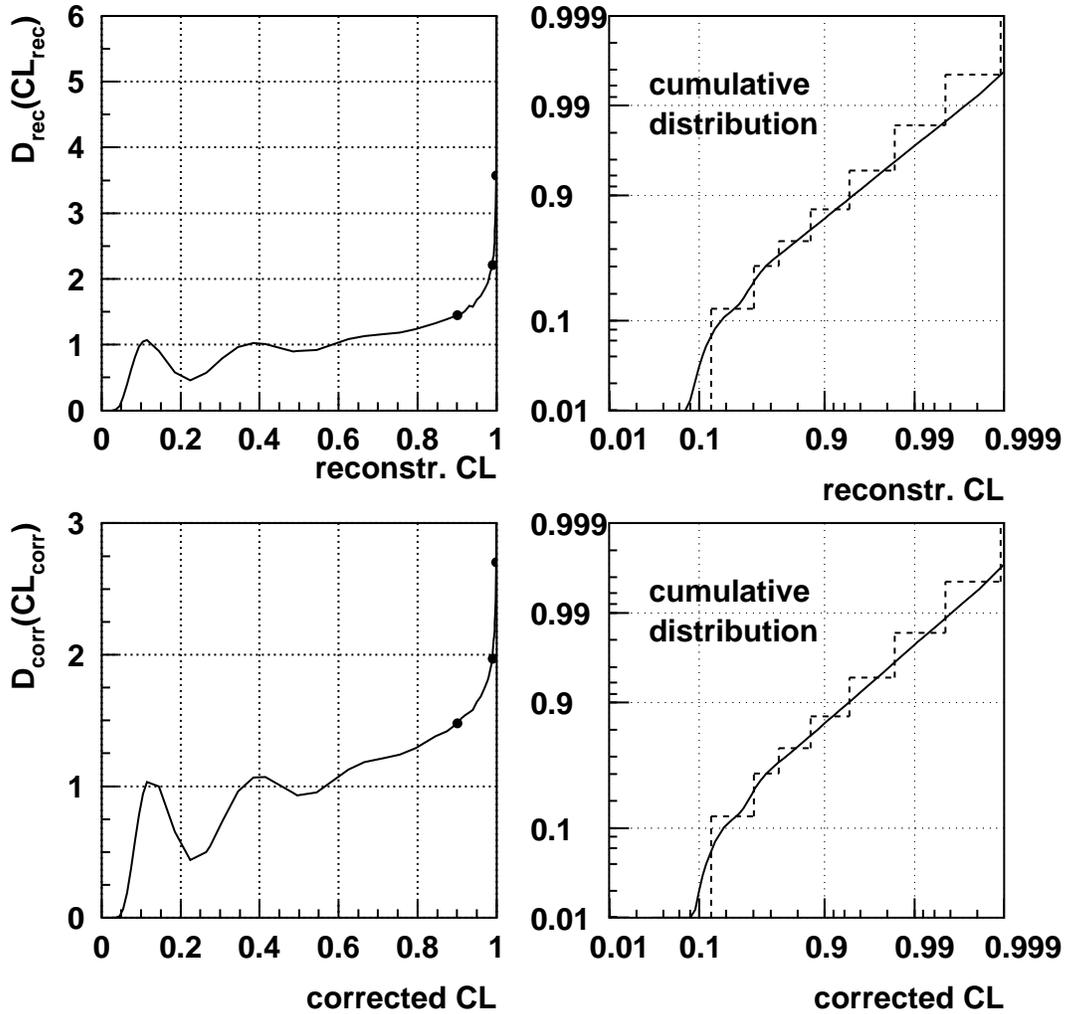,height=15cm}
     }
\caption {{\it Spectra of confidence levels for a Poisson distribution
with $n_0$ = 2 and a systematic error of 20\%, as reconstructed by many 
observers.
Lower (upper) part: The confidence levels are corrected (not
corrected) for systematic errors.
Left: Differential distributions. The dots mark the results for the
reconstructed confidence levels 90\%, 99\% and 99.9\%.
Right: Cumulative distributions. The step functions show the true
original cumulative distribution of $CL$.}}
\label{pois}
\end{center}
\end{figure}

\begin{figure}[htb]
\begin{center}
\mbox{
\epsfig{file=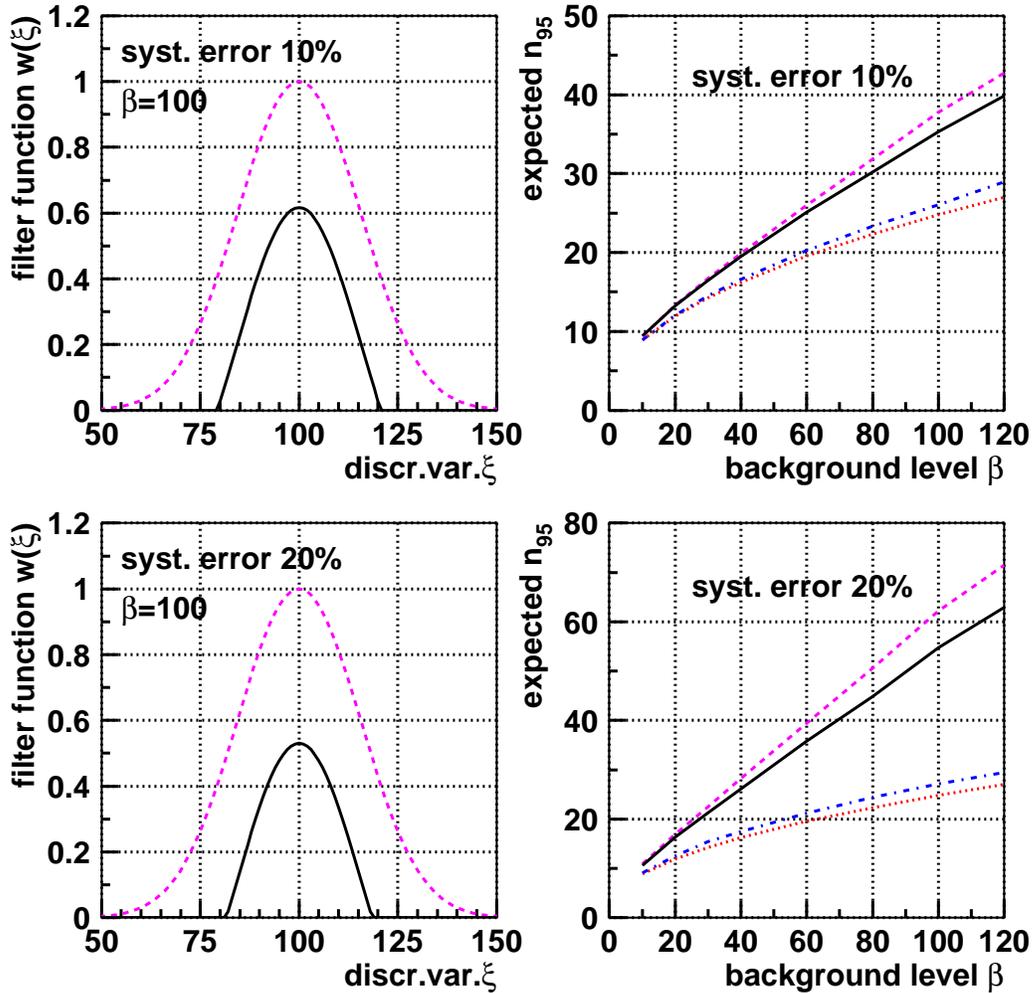,height=15cm}
}
\caption{{\it Expected upper rate limits for a non-existing
Gaussian signal over a constant background.
Left column: Weight functions. Dashed curves: weighting based on 
statistical errors only. Full curves: systematic errors included in
the weights. The ordinate scale is arbitrary.
Right column:
95\% limits as a function of the background level.
The lower two curves give the limits based on statistical
errors only. The dotted (dash-dotted) lines 
correspond to the dashed (full) curves on the 
left hand side. Upper curves: Systematic errors are taken
into account.
}}
\label{wsysopt}
\end{center}
\end{figure} 

\end{document}